\documentclass[12pt,reqno]{amsart}
\pdfoutput=1
\usepackage[headings]{fullpage}
\usepackage{amssymb,bbm}
\usepackage{epstopdf}
\usepackage{graphicx}
\usepackage{texdraw}
\usepackage{url}
\usepackage[all]{xy}
\usepackage{pb-diagram}
\usepackage{float}   
\usepackage[bookmarks=true,%
    colorlinks=true,%
    linkcolor=blue,%
    citecolor=blue,%
    filecolor=blue,%
    menucolor=blue,%
    urlcolor=blue,%
    breaklinks=true]{hyperref}

\newtheorem{theorem}{Theorem}

\theoremstyle{definition}

\newtheorem{conjecture}[theorem]{Conjecture}

\newtheorem{example}[theorem]{Example}

\def\BN{\mathbbm N}
\def\BZ{\mathbbm Z}
\def\BQ{\mathbbm Q}
\def\BR{\mathbbm R}
\def\BC{\mathbbm C}

\def\calI{\mathcal I}

\def\CK{\mathcal K}
\def\CI{\mathcal I}
\def\IZ{\mathbb{Z}}

\def\calP{\mathcal P}
\def\calI{\mathcal I}
\def\s{\sigma}

\def\SL{\mathrm{SL}}

\def\Ind{\mathrm{Ind}}

\def\tq{\tilde{q}}
\def\Res{\mathrm{Res}}

\def\={\;=\;}
\def\+{\,+\,}
\def\-{\,-\,}

\def\be{\begin{equation}}
\def\ee{\end{equation}}

\def\J{\mathbf J}

\def\Ind{\mathrm{Ind}}
\def\rot{\mathrm{rot}}
\def\ve{\varepsilon}
\def\bb{\mathsf{b}}
\def\ri{\mathrm{i}}
\def\rd{\mathrm{d}}
\def\wt{\widetilde}
\def\tx{\tilde{x}}
\def\vev#1{\langle #1 \rangle}
\def\re{{\rm e}}
\def\ri{{\rm i}}
\def\rd{{\rm d}}
\newcommand{\figref}[1]{Fig.~\protect\ref{#1}}
\def\SU{\mathrm{SU}}

\def\mc{\mathcal}
\def\md{\mathbf}
\def\mf{\mathfrak}
\def\ms{\mathsf}

\def\cO{O}

\newcommand{\nn}{\nonumber \\}

\begin{document}
\title[The resurgent structure of quantum knot invariants]{The
  resurgent structure of quantum knot invariants}
\author{Stavros Garoufalidis}
\address{
  International Center for Mathematics, Department of Mathematics \\
  Southern University of Science and Technology \\
  Shenzhen, China \newline
  {\tt \url{http://people.mpim-bonn.mpg.de/stavros}}}
\email{stavros@mpim-bonn.mpg.de}
\author{Jie Gu}
\address{D\'epartement de Physique Th\'eorique, Universit\'e de Gen\`eve \\
Universit\'e de Gen\`eve, 1211 Gen\`eve 4, Switzerland}
\email{jie.gu@unige.ch}
\author{Marcos Mari\~no}
\address{Section de Math\'ematiques et D\'epartement de Physique Th\'eorique\\
Universit\'e de Gen\`eve, 1211 Gen\`eve 4, Switzerland }
\email{marcos.marino@unige.ch}



\date{29 November 2020}

\begin{abstract}
  The asymptotic expansion of quantum knot invariants in complex
  Chern-Simons theory gives rise to factorially divergent formal power
  series. We conjecture that these series are resurgent functions
  whose Stokes automorphism is given by a pair of matrices of
  $q$-series with integer coefficients, which are determined
  explicitly by the fundamental solutions of a pair of linear
  $q$-difference equations. We further conjecture that for a
  hyperbolic knot, a distinguished entry of those matrices equals to
  the Dimofte-Gaiotto-Gukov 3D-index, and thus is given by a counting
  of BPS states. We illustrate our conjectures explicitly by matching
  theoretically and numerically computed integers for the cases of the
  $4_1$ and the $5_2$ knots.
\end{abstract}

\maketitle

{\footnotesize
\tableofcontents
}


\section{Introduction}
\label{sec.intro}

\subsection{Asymptotic expansions in perturbative quantum
  field theory}
  
Perturbative expansions in quantum field theories are often
mathematically defined but typically lead to factorially divergent
formal power series. Important examples are the perturbative
expansions of the partition function of 3-dimensional manifolds (with,
or without boundary) in complex Chern-Simons theory with arbitrary
gauge group. For instance, in~\cite{GLM} it was shown that the LMO
invariant of a 3-manifold (which is the perturbative expansion of the
Witten-Reshetikhin-Turaev invariant at the trivial flat collection and
arbitrary gauge group), is a Gevrey-1 formal power series, that is a
formal power series whose $n$th coefficient is bounded by $n! C^n$ for
some positive constant $C$.  Said differently, these perturbative
expansions have Borel transforms which are germs of holomorphic
functions at the origin.

In~\cite{Ga:resurgence} it was conjectured that the perturbative
expansions in complex Chern-Simons theory are resurgent functions, and
more precisely that they have analytic continuation as multivalued
functions in the complex plane minus a discrete and computable set of
points, placed in finitely many vertical lines in the complex
plane. These vertical lines are formed by an infinite towers of
singularities, with a $2 \pi \ri$ periodicity. The position of these
singularities is dictated by the values of the complex Chern-Simons
function (a $\BC/2 \pi {\rm i} \BZ$-valued function) on the set of
flat connections.

\subsection{Resurgence in complex Chern-Simons theory}

In what follows, we will identify the partition function of complex
Chern-Simons theory with the state integral of Andersen-Kashaev
in~\cite{AK}, following the ideas of Hikami, Dimofte et
al~\cite{Hikami,DGLZ}.  Although this identification has not been
derived from first principles, it turns out to have a number of
startling consequences. We will focus on manifolds of the form
$M= \mathbb{S}^3 \backslash \mathcal{K}$, where $\CK$ is a hyperbolic
knot.

Our goal is to give an explicit description of the resurgent structure
of the formal power series of perturbative Chern-Simons theory in
terms of a fundamental solution of a pair of linear $q$-difference
equation and a matrix of integers. We will describe the general story
first, and illustrate it with concrete examples later.

We will denote by $\calP$ the set of critical points of the complex
Chern-Simons action and by $\s$ a typical critical point.  Given our
identification of Chern-Simons theory with state integrals, it turns
out that the set $\calP$ coincides with the set of critical points of
the integrand of the state integral, an effectively computable set of
algebraic numbers. The critical values of the complex Chern-Simons
function are labeled by $\s$ and an integer $\mu$ (often called
``multicovering''):
\begin{equation}
  \label{actions}
  \text{CS}(\s;\mu) =
  {V(\s) \over 2\pi} - 2\pi {\ri} \mu, \qquad \mu \in \BZ. 
\end{equation}
We conjecture that the corresponding transseries $ \Phi_{\s, \mu}$
satisfy the translation invariance property
\begin{equation}
  \label{Phiqt}
  \Phi_{\s,\mu}(\tau) = \tq^\mu \, \Phi_{\s}(\tau), \qquad
  \tq = {\rm e}^{-2\pi {\rm i}/\tau}, 
\end{equation}
where $\Phi_{\s}(\tau)$ is the conventional asymptotic expansion of
the state integral around the saddle point $\s$. It has the form
\begin{equation}
  \label{phi-ex}
  \Phi_{\s}(\tau)=
  \exp\left( { V(\s) \over 2\pi \tau} \right) \varphi_\s (\tau),
  \qquad \varphi_\s (\tau) \in \BC[[\tau]]. 
\end{equation}
As a consequence, all the Stokes automorphisms acting on
$\Phi_{\s,\mu}$ are packaged in two Stokes automorphism matrices
$\ms{S}^{+}(q)$, $\ms{S}^-(q^{-1})$, which are matrices of $q$- and
$q^{-1}$-series respectively, and each of which encodes Stokes
automorphism across a half-plane.  Their detailed definition is given
in Section~\ref{sc:res}.


An important feature of state integrals is that they depend on
additional parameters and this leads to a system of a pair of linear
$q$-difference equations, one in the upper half-plane and another in
the lower half-plane~\cite{GK:qseries}. In our examples, these linear
$q$-difference equations have explicit sets of fundamental
solutions. We conjecture that

\begin{conjecture}
  \label{conj.1}
  \rm{(a)} $\mathsf{S}^{\pm}(q)$ are bilinear functions of two
  fundamental solutions of the pair of linear $q$-difference
  equations. \newline 
  \rm{(b)} $\mathsf{S}^{\pm}(q)$ satisfy the inversion relation
  \begin{equation}
    \label{Ssym}
    \ms{S}^+(q)^T \ms{S}^-(q) = \md{1} \,.
  \end{equation}
  \rm{(c)} $\mathsf{S}^{\pm}(q)$ are uniquely determined by
  $\mathsf{S}^+(0)$, $\mathsf{S}^-(0)$ and a pair of fundamental
  solutions to the pair of linear $q$-difference equations.
\end{conjecture}

The above matrices $\mathsf{S}^{\pm}$ uniquely determine the collection of
transseries $\Phi_{\s,\mu}(\tau)$ for all $(\s,\mu)$ via an abstract
Riemann-Hilbert correspondence first pointed out in \cite{voros} and
developed recently in \cite{gmn,ims,ks-ar}. Note that this
transcendental correspondence converts the difficult problem of
computing coefficients of $\Phi_{\s,\mu}(\tau)$ (typically, one can
not compute more than a couple of hundred coefficients) into the much
easier problem of computing fundamental solutions of linear
$q$-difference equations, up to a matrix of unknown integers.

Given a hyperbolic knot $\CK$ there is a distinguished critical point
$\s_1$ (the geometric representation, corresponding to the complete
hyperbolic structure), and in that case we conjecture a precise
relation between the entry $\mathsf{S}^+_{\s_1,\s_1}(q)$ of the matrix
$\ms{S}^+(q)$ and the (rotated) 3D-index of
Dimofte-Gaiotto-Gukov~\cite{DGG1,DGG2}.

\begin{conjecture}
  \label{conj.2}
  We have:
  \begin{equation}
    \label{Sind}
    \mathsf{S}^+_{\s_1\s_1}(q) =  \,\Ind_\CK^\rot(q)\,.
  \end{equation}
\end{conjecture}

We recall that the 3D-index $\CI_\CK (m,e)(q)$ associated to a knot
$\CK$ is labeled by two integers $(m,e)$. It counts BPS states in a
three-dimensional, $\mathcal{N}=2$ supersymmetric theory $T_M$ which
can be associated to the manifold $M=\mathbb{S}^3\backslash \CK$
\cite{DGG1}.  The rotated index is then given by
\begin{equation}
  \Ind_\CK^\rot(q)=\sum_{e \in \IZ} \CI_\CK (0,e)(q). 
\end{equation}
The relation in \eqref{Sind} between the resurgent structure of
complex Chern--Simons theory and a counting of BPS states in the
corresponding supersymmetric theory was anticipated in
\cite{mm-s2019,ggm}.

We emphasize that although the state integrals and their perturbation
theory are well-defined, the above picture is largely conjectural.
However, it fits well with the work of Kontsevich-Soibelman~\cite{KS},
as well as with a lecture of Kontsevich on June 30, 2020
\cite{kontsevich-talk}, and it paves the way for a deeper
understanding of the topological/physical meaning of the integers
appearing in $\mathsf{S}^{\pm}$.


We should point out that the above theory in fact has little to do
with knots and 3-manifolds and complex Chern-Simons theory, and little
to do with the Bloch group, but appears to be part of a larger
combinatorial structure. This is apparent in the data needed to define
the formal power series of~\cite{DG,DG2} as well as the data needed to
define $q$-hypergeometric Nahm sums (and thus their asymptotic
expansion at roots of unity~\cite{GZ:asymptotics}) and the data needed
to define state integrals~\cite{GK:qseries}. This combinatorial
structure is sometimes called a $K_2$-Lagrangian, or an extended
symplectic group.

We will illustrate the above conjectures concretely for the invariants
of the two simplest hyperbolic knots, the $4_1$ and the $5_2$
knots. Some aspects of the resurgent structure of complex
Chern--Simons theory for the $4_1$ knot were studied in
\cite{gmp,gh-res}, but they focused on the ``classical" transseries
$\Phi_{\s}(\tau)$ (i.e. they didn't address the resurgent structure of
the tower of singularities). The resurgent problem in the case of
compact $\SU(2)$ Chern--Simons theory was addressed in \cite{CG, gmp},
where complete towers of Stokes constants were explicitly computed for
some Seifert three-manifolds.

This paper is, in a sense, a sequel to~\cite{GZ:kashaev}
and~\cite{GZ:qseries} which the reader can consult for further
information, motivation, historical presentation, as well as for the
connection with the asymptotics of the Kashaev invariant and with the
quantum modularity conjecture.

Note that our notation $\varphi_\s(\tau)$ from Equation~\eqref{phi-ex}
corresponds to the notation $\Phi^{(\s)}_0(2\pi \ri x)$
of~\cite{GZ:kashaev}.  In particular, the coefficient of $\tau^n$ in
$\varphi_\s(\tau)$ is (up to multiplication by an eighth root of unity
and the square root of an nonzero element of $F_\s$) in
$(2\pi \ri)^n F_\s$, where $F_\s$ is the trace field of $\s$.


\section{The equation $(1-x)(1-x^{-1})=1$ and the $4_1$ knot}
\label{sec.41}


The state integral of the $4_1$ knot is given by
Equation~\eqref{eq.statelm} below with $(A,B)=(1,2)$ and
$\mu=\lambda=0$. The critical points of the integrand are solutions of
the algebraic equation
\begin{equation}
  \label{41xi}
  (1-x)(1-x^{-1})=1.
\end{equation}
The latter has two solutions $\xi_1= {\rm e}^{2 \pi \ri/6}$ and
$\xi_2= \re^{-2 \pi \ri/6}$ which lie in the number field
$\BQ(\sqrt{-3})$, the trace field of the $4_1$ knot. The corresponding
series $\Phi_{\s_j}(\tau)$ satisfy the relation
$\Phi_{\s_2}(\tau) = {\rm i} \Phi_{\s_1}(-\tau)$ and the first few
terms of
$\varphi_{\s_1}(\tau/(2\pi \ri)) \in 3^{-1/4} \BQ(\sqrt{-3})[[\tau]]$
are given by
\begin{equation}
  \label{varphi41}
  \varphi_{\s_1}\left(\frac{\tau}{2\pi \ri}\right) =
  \frac{1}{\sqrt[4]3}\, 
  \Bigl(1 \+ \frac{11\tau}{72\sqrt{-3}}
  \+ \frac{697\tau^2}{2\,(72\sqrt{-3})^2}
  \+ \frac{724351\tau^3}{30\,(72\sqrt{-3})^3} \+\cdots\Bigr)\,. 
\end{equation}
The exponent in \eqref{phi-ex} involves the hyperbolic volume of the
$4_1$ knot complement
\begin{equation}
  \label{volume-41}
  V(\s_1)= V=2 {\rm Im}\, {\rm Li}_2(\re^{\ri \pi/3})= 2.029883\dots \,.
\end{equation}
The two series $\Phi_{\s_{j}}(\tau)$ for $j=1,2$ form a vector
\begin{equation}
  \label{phivec}
  \Phi(\tau)=
  \begin{pmatrix} \Phi_{\s_1}(\tau)\\ \Phi_{\s_2}(\tau) \end{pmatrix}
\end{equation}
that also appears in the refined quantum modularity
conjecture~\cite{GZ:kashaev}. $\Phi(\tau)$ is the vector of series
whose description in Borel plane we wish to give.

Consider the linear $q$-difference equation

\begin{equation}
  \label{41qdiff}
  y_{m+1}(q) -(2-q^m) y_m(q) + y_{m-1}(q)=0 \qquad (m \in \BZ) \,.
\end{equation}
It has a fundamental solution set given by the columns of the
following matrix
\begin{equation}
  \label{Gfund}
  W_m(q) =
  \begin{pmatrix} G^0_m(q) & G^1_m(q) \\ G^0_{m+1}(q) & G^1_{m+1}(q)
  \end{pmatrix},
\end{equation}  
where $G^0_m(q)$ and $G^1_m(q)$ are defined by
\begin{subequations}
  \begin{align}
\label{gm}
    G^0_m(q)
    &=\sum_{n=0}^\infty (-1)^n \frac{q^{n(n+1)/2+m n}}{(q;q)_n^2}
    \\
\label{Gm}
    G^1_m(q)
    &=\sum_{n=0}^\infty (-1)^n \frac{q^{n(n+1)/2+m n}}{(q;q)_n^2}
      \left(2m+ E_1(q) + 2 \sum_{j=1}^n \frac{1+q^j}{1-q^j} \right) \,,
\end{align}
\end{subequations}
and $E_1(q)=1-4\sum_{n=1}^\infty q^n/(1-q^n)$ is the Eisenstein series. 

It is easy to see that $G^0_m$ satisfies~\eqref{41qdiff}. Indeed,
$G^0_m(q)=\sum_{n=0}^\infty a_{n,-m}(q)^{-1}$ where $a_{n,m}(q)$ is
given in~\eqref{anm41} below. Equations~\eqref{ap1}-\eqref{ap3},
applied to $a_{n,-m}(q)^{-1}$ conclude the result. A similar proof
applies for $G^1_m$. Another way to do so is to use the state
integral~\eqref{eq.statelm} (with $(A,B)=(1,2)$) and show that the
latter satisfies the linear $q$-difference equation~\eqref{41qdiff} in
two ways, one with respect to the variable $\lambda$ and another with
respect to the variable $\mu$.

The fundamental solution $W_m(q)$ satisfies
\begin{equation}
  \label{det41}
  \det(W_m(q))=2
\end{equation}
and the symmetry
\begin{equation}
  \label{W41inv}
  W_m(q^{-1}) = W_{-m}(q)
  \begin{pmatrix} 1 & 0 \\ 0 & -1 \end{pmatrix} \,,
\end{equation}
and the orthogonality
\begin{equation}
  \label{WWT41b}
  \frac{1}{2} W_m(q)
  \begin{pmatrix} 0 & 1 \\ -1 & 0 \end{pmatrix}
  W_{m}(q)^T =
  \begin{pmatrix} 0 & 1 \\ -1 & 0 \end{pmatrix} \,.
\end{equation}
for all integers $m$, as well as
\begin{equation}
  \label{WWT41}
  \frac{1}{2} W_m(q)
  \begin{pmatrix} 0 & 1 \\ -1 & 0 \end{pmatrix}
  W_{\ell}(q)^T \in \SL(2,\BZ[q,1/q]) 
\end{equation}
for all integers $m, \ell$. The $\mathsf{S}$ matrix is given by 
\begin{subequations}
  \begin{align}
    \label{S41p}
    \mathsf{S}^+(q)
    &={1\over 2}
      \begin{pmatrix} 0 & 1 \\ 1 & 1 \end{pmatrix}
       W_{-1}(q) \begin{pmatrix} 0 & 1 \\ 1 & 0 \end{pmatrix}W_{-1}(q)^T
  \begin{pmatrix} 0 & -1 \\ 1 & 2 \end{pmatrix},\\
  \label{S41m}
  \mathsf{S}^-(q) &= {1\over 2}
  \begin{pmatrix} -1 & -1 \\ 0 & 1 \end{pmatrix} W_{-1}(q)
  \begin{pmatrix} 0 & 1 \\ 1 & 0 \end{pmatrix}W_{-1}(q)^T
  \begin{pmatrix} 1 & 0 \\ -2 & 1 \end{pmatrix} \,.
\end{align}
\end{subequations}
The above matrix $\ms{S}$ satisfies Equation~\eqref{Ssym}.


\section{The equation $x^{-2}(1-x)^3=1$ and the $5_2$ knot}
\label{sec.52}

The state integral of the $5_2$ knot is given by
Equation~\eqref{eq.statelm} below with $(A,B)=(2,3)$ and
$\mu=\lambda=0$. The critical points of the integrand are solutions of
the algebraic equation
\begin{equation}
  \label{52xi}
  x^{-2}(1-x)^3=1 \,.
\end{equation}
The above equation (which defines a cubic field of discriminant $-23$,
the trace field of the $5_2$ knot) has three solutions
$\xi_1=0.78492+1.30714\dots \ri$, $\xi_2=0.78492-1.30714\dots \ri$ and
$\xi_3=0.43016\dots $, corresponding to the geometric representation,
its conjugate and the real representation.  The corresponding series
$\Phi_{\s_j}(\tau)$ satisfy
$\Phi_{\sigma_2}(\tau) = -\ri \overline{\Phi}_{\sigma_1}(-\tau)$ and
the first few terms of $\phi_{\s_j}(\tau/(2\pi \ri))$ are given by
\begin{align}
  \notag
  \varphi_{\s_j}\left(\frac{\tau}{2\pi \ri}\right)
  &=
    \left(\frac{-3\xi_j^2+3\xi_j-2}{23}\right)^{1/4}
    \left(1+\frac{33\xi_j^2+242\xi_j-245}{2^2\cdot 23^2}\tau
    +\frac{100250\xi_j^2-12643\xi_j+2732}{2^5\cdot 23^3}\tau^2\right.
  \\
  \notag
  &+\frac{-50198891\xi_j^2+35443870\xi_j-79016748}
    {2^7\cdot 3\cdot 5\cdot 23^5}\tau^3
  \\
    \label{varphi52}
  &\left.+\frac{-3809943572\xi_j^2+1861268771\xi_j+1015686665}
    {2^{11}\cdot 3\cdot 5 \cdot 23^6}\tau^4+\dots
    \right) \,.
\end{align}
The exponent in \eqref{phi-ex} involves 
\begin{equation}
  V(\sigma_1) = 2.821\ldots+ 1.379\ldots \ri,\;
  V(\sigma_2) = -2.821\ldots+ 1.379\ldots \ri,\;
  V(\sigma_3) = -2.758\ldots 
  \label{eq:V_52}
\end{equation}
where $\Re\,V(\sigma_1)$ is the hyperbolic volume of the $5_2$ knot
complement, and $\Im\,V(\sigma_1) = \Im\,V(\sigma_2)$ the Chern-Simons
action. The three series $\Phi_{\sigma_j}(\tau)$ for $j=1,2,3$ form a
vector
\begin{equation}
  \Phi(\tau) = \left(
    \begin{array}{c}
      \Phi_{\sigma_1}(\tau)\\
      \Phi_{\sigma_2}(\tau)\\
      \Phi_{\sigma_3}(\tau)
    \end{array}
  \right).
  \label{eq:phivec_52}
\end{equation}

Consider the linear $q$-difference equation 
\begin{equation}
  \label{52qdiff}
  y_m(q)-3y_{m+1}(q)+(3-q^{2+m})y_{m+2}(q) - y_{m+3}(q) = 0.
\end{equation}
The above equation has a fundamental solution set given by the
columns of the following matrix
\begin{equation}
  \label{52fund}
  W_m(q) =
  \begin{cases}
    W_m^+(q),\quad &|q|<1,\\
    \begin{pmatrix}
      0&0&1\\
      0&1&0\\
      1&0&0
    \end{pmatrix}
    W_{-m-2}^-(q^{-1})
    \begin{pmatrix}
      1&0&0\\
      0&-1&0\\
      0&0&1
    \end{pmatrix},\quad &|q|>1.
  \end{cases}
\end{equation}
where the matrices $W_m^\ve(q)$ with $\ve=\pm$ are respectively
\begin{equation}
  \label{52fund1}
  W^\ve_m(q) =
    \begin{pmatrix}
      H^{\ve}_{0,m}(q) & H^{\ve}_{1,m}(q) & H^{\ve}_{2,m}(q)       \\
      H^{\ve}_{0,m+1}(q) & H^{\ve}_{1,m+1}(q) & H^{\ve}_{2,m+1}(q) \\
      H^{\ve}_{0,m+2}(q) & H^{\ve}_{1,m+2}(q) & H^{\ve}_{2,m+2}(q)
    \end{pmatrix} 
\end{equation}
and $H^\ve_{j,m}(q)$ are given in Appendix~\ref{sec.A52} for $j=0,1,2$
and $m \in \BZ$.


The fundamental solutions satisfy
\begin{equation}
  \label{det52}
  \det(W_m(q))=2,
\end{equation}
for all integers $m$ as well as
\begin{equation}
  \label{WWT52}
  \frac{1}{2} W_m(q)
  \begin{pmatrix} 0&0&1 \\ 0&2&0 \\ 1&0&0 \end{pmatrix}
  W_{\ell}(q^{-1})^T \in \SL(3,\BZ[q,1/q]) 
\end{equation}
for all integers $m, \ell$. In particular, we have:
\begin{equation}
  \label{WWT52b}
  \frac{1}{2} W_{m-1}(q)
  \begin{pmatrix} 0&0&1 \\ 0&2&0 \\ 1&0&0 \end{pmatrix}
  W_{-m-1}(q^{-1})^T =
  \begin{pmatrix} 1&0&0 \\ 0&0&1 \\ 0&1&3-q^{m} \end{pmatrix} \,.
\end{equation}
The $\mathsf{S}$ matrix is given by 
\begin{subequations}
\begin{align}
  \label{S52p}
  \mathsf{S}^+(q) &= \frac{1}{2}
      \begin{pmatrix}
        0&1&0\\
        0&1&1\\
        -1&0&0
      \end{pmatrix} W_{-1}(q^{-1})
               \begin{pmatrix}
                 0&0&1\\
                 0&-2&0\\
                 1&0&0
               \end{pmatrix} W_{-1}(q)^T
               \begin{pmatrix}
                 0&0&-1\\
                 1&1&0\\
                 0&1&0
               \end{pmatrix},\\
  \label{S52m}
  \mathsf{S}^-(q) &= \frac{1}{2}
      \begin{pmatrix}
        0&3&-1\\
        0&-1&0\\
        1&0&0
      \end{pmatrix}
              W_{-1}(q)
              \begin{pmatrix}
                 0&0&1\\
                 0&-2&0\\
                 1&0&0
               \end{pmatrix} W_{-1}(q^{-1})^T
              \begin{pmatrix}
                0&0&1\\
                3&-1&0\\
                -1&0&0
              \end{pmatrix}\,.
\end{align}
\end{subequations}
The above matrix $\ms{S}$ satisfies Equation~\eqref{Ssym}. A proof is
given in Appendix~\ref{sub.qsym}.


\section{Descendants}
\label{sec.d}

A key aspect of our study of asymptotic series are linear
$q$-difference equations which are satisfied for their
descendants. This elementary idea leads to descendants of the Kashaev
invariant (studied extensively in~\cite{GZ:kashaev}), of asymptotic
series (ibid), of $q$-series as well as of state integrals. In this
section we review in detail the story of descendants (or ancestors, as
the case may be).

\subsection{The Kashaev invariant and its descendants}
\label{sub.k41}

The series $\Phi_{\s_1} (\tau)$ appearing in the saddle-point
expansion of the state integral appeared originally in the asymptotic
expansion of the Kashaev invariant \cite{K95}. In the case of the
$4_1$ knot, the Kashaev invariant is given by
\begin{equation}
\label{J41}
\J^{(4_1)}(q) = \sum_{n=0}^\infty (q;q)_n (q^{-1};q^{-1})_n \,.
\end{equation}
The above expression can be evaluated when $q$ is a root of unity. The
Volume Conjecture of Kashaev~\cite{kas-volume} (and its extension to all orders \cite{gukov}) 
asserts that $\J^{(4_1)}(\re^{2 \pi \ri/N})$ has
an asymptotic expansion for $N$ large of the form
\begin{equation}
\label{J41Phi}
\J^{(4_1)}(\re^{2 \pi \ri/N}) \sim N^{3/2}\,
\Phi_{\s_1}\left(\frac{1}{N}\right). 
\end{equation}


We now explain a relation discovered in~\cite{GZ:kashaev} 
between the formula for the Kashaev invariant~\eqref{J41} and the
algebraic equation~\eqref{41xi}.

Following~\cite{GZ:kashaev}, we define the descendants
$\J^{(4_1)}_m(q)$ of the Kashaev invariant of the $4_1$ knot by
\begin{equation}
\label{J41m}
\J^{(4_1)}_m(q) = \sum_{n=0}^\infty (q;q)_n (q^{-1};q^{-1})_n \, q^{m n},
\qquad (m \in \BZ) \,.
\end{equation}
Then, the sequence $\J^{(4_1)}_m(q)$ is a solution to a linear $q$-difference
equation
\begin{equation}
\label{qdiffJ}
q^{m+1} \J_m(q) + (1-2q^m)\J_m(q) + q^{m-1} \J_{m-1}(q) =1,
\qquad (m \in \BZ) \,.
\end{equation}
This can be seen as follows: let
\begin{equation}
\label{anm41}
a_{n,m}(q)=
(q;q)_n (q^{-1};q^{-1})_n q^{m n}
\end{equation}
denote the summand of~\eqref{J41m}. It follows that
\begin{subequations}
  \begin{align}
    \label{ap1}
    a_{n+1,m}(q) &= (1-q^{-n-1})(1-q^{n+1}) \,a_{n,m}(q) \\
    \label{ap2}
    &= q^m(2-q^{n+1}-q^{-n-1}) \, a_{n,m}(q) \\
    \label{ap3}
  &= q^m(2-q \, a_{n,m+1}(q)-q^{-1} a_{n,m-1}(q)) \,.
\end{align}
\end{subequations}
Summing over $n \geq 0$ and taking into account the boundary term
$a_{0,m}(q)=1$ on the left hand side of the above equation concludes the
proof of Equation~\eqref{qdiffJ}.

Using the operators $E$ and $Q$ that act on sequences $(y_m)$ by
\begin{equation}
\label{EQ}
(Ey)_m=y_{m+1}, \qquad (Qy)_m=q^m y_m, \qquad EQ=qQE
\end{equation}
it follows that we can write~\eqref{qdiffJ} in the form
\begin{equation}
\label{qdiffJ2}
( Q(1-qE)(1-q^{-1}E^{-1}) - I) \J_m(q) = 1 \,.
\end{equation}
The homogeneous part of the above operator can be obtained by
replacing $x$ in the left hand side of Equation~\eqref{41xi} $x$ by
$qE$, and replacing the right hand side of Equation~\eqref{41xi} by
$Q^{-1}$.

\subsection{The $q$-series $(G^0_0,G^1_0)$ and their
  descendants}
\label{sub.gG41}

We now discuss an appearance of the formal power series $\Phi(\tau)$
in the radial asymptotics of some $q$-series, following~\cite{GZ:qseries}.

By $q$-series we mean formal Laurent series in a variable $q$ with
integer coefficients, i.e., elements of $\BZ((q))$.  All the
$q$-series below will define holomorphic functions in the punctured
unit disk with (perhaps) a pole at the origin.  We now recall how the
radial asymptotics of the $q$-series $(G^0_0,G^1_0)$ is given by
$\Phi(\tau)$. The first series $G^0_0(q)$ was found quite by accident
to have radial asymptotics expressed in terms of the series
$\Phi(\tau)$~\cite{GZ:qseries}, whereas the second series was found
systematically by expressing the state integral invariant of the $4_1$
knot in terms of products of $q$-series and
$\tq$-series~\cite{GK:qseries}.

Below, we will use capital letters for $q$-series and small letters
for the corresponding functions on the upper half-plane, e.g.,
$g^0_m(\tau)=G^0_m(q)$ for $q=\re^{2\pi
  \ri\tau}$. In~\cite{GZ:qseries} it was observed that we have an
asymptotic expansion

\begin{equation}
\label{gGtau}
\begin{pmatrix} \frac{1}{\sqrt{\tau}} g^0_0(\tau) \\
  \sqrt{\tau} g^1_0(\tau)
\end{pmatrix} \sim \begin{pmatrix} 1 & -1 \\ 1 & 1 \end{pmatrix}
  \Phi( \tau) 
\end{equation}
to all orders in $\tau$, as $\tau$ tends to $0$ along a ray in the
first quadrant of the upper half-plane. The above asymptotic expansion
requires some explanation since on a fixed ray in the first quadrant,
$\Phi_{\s_1}( \tau)$ is exponentially larger than
$\Phi_{\s_2}(\tau)$. Nonetheless, the asymptotic
expansion~\eqref{gGtau} makes sense theoretically and computationally
if we use a refined optimal truncation explained in detail
in~\cite{GZ:kashaev} and applied in~\cite{GZ:qseries}. The numerical
computations of~\cite{GZ:qseries} hinted that the matrix in
Equation~\eqref{gGtau} is the constant term of a matrix of $\tq$
series.

Given the definition of $G^0_0(q)$ and $G^1_0(q)$ from~\cite{GZ:qseries}
and~\cite{GK:qseries}, it was relatively straightforward to add the
variable $q^{m n}$ and arrive to formulas~\eqref{gm} and~\eqref{Gm}
which define the descendants of the pair $(G^0_0(q),G^1_0(q))$.

\subsection{The state integral and its descendants}
\label{sub.state}

In this section we recall the definition of state integrals and some
of their basic properties. State integrals are multidimensional
integrals whose integrand is a product of Faddeev's quantum
dilogarithm function $\Phi_{\bb}$ (whose definition we will not need
and may be found in~\cite{Faddeev,FK-QDL}) times an exponential of a
quadratic and linear form. Here
$\tau=\bb^2 \in \BC':=\BC\setminus(-\infty,0]$, thus even if the
integrand contains no free variables, a state integral is always a
holomorphic function of $\tau$.

State integrals have two key properties:
\begin{itemize}
\item[(a)] they define holomorphic functions in the complex cut plane
  $\BC'$.
\item[(b)] they can be expressed bilinearly in the upper and in the
  lower half-plane in terms of products of $q$-series and
  $\tq$-series.
\end{itemize}
For a detailed discussion of state integrals and numerous example, see
for instance~\cite{Beem} and also~\cite{AK} and~\cite{GK:qseries}.

In this section we introduce a descendant version of the one
dimensional state integrals of~\cite{GK:qseries} that satisfies the
above properties. In this section we will use the notation
from~\cite{GK:qseries}.  Consider the state integral

\begin{equation}
\label{eq.statelm}
  \calI_{A,B,\lambda,\mu}(\bb) =
    \int_{\BR+\ri\epsilon} \Phi_{\bb}(x)^B \re^{-A\pi\ri x^2 +
      2\pi(\lambda \bb - \mu \bb^{-1})x}\rd x,\qquad
     (\lambda,\mu\in \BZ)
   \end{equation}
   where $A$ and $B$ are integers and $B > A > 0$. Under these
  assumptions, it follows that the integrand is exponentially decaying
  at infinity and the integral is absolutely convergent and defines
  a holomorphic
  function of $\tau = \bb^2 \in \BC'$.
  Below, we will use the notation
  $\phi(w,\delta_\bullet)$, $\tilde{\phi}(w,\tilde{\delta}_\bullet)$
  and 
  \begin{equation}
    \vev{F(q,x)} = F(q,1)
  \end{equation}
  from~\cite{GK:qseries}.
  
  \begin{theorem}
    \label{thm.state}
    Fix integers $A$ and $B$ with $B > A > 0$ and integers $\lambda$ and
    $\mu$. For all $\tau$ with $\Im(\tau)>0$, we have:
\begin{equation}
\label{thm1}
    \calI_{A,B,\lambda,\mu}(\bb) =
    (-1)^{\lambda-\mu} q^{\frac{\lambda}{2}}\tq^{\frac{\mu}{2}}
    \left(\frac{\tq}{q}\right)^{\frac{B-3A}{24}}
    \re^{\pi\ri\frac{B+2(A+1)}{4}}
    \vev{P_{A,B,\lambda,\mu}
      \left(F_{A,B,\lambda}(q,x)\wt{F}_{A,B,\mu}(\tq,\tx)\right)}
  \end{equation}
where the operator $P_{A,B,\lambda,\mu}$ is given by
\begin{equation}
  \label{Plm}
    P_{A,B,\lambda,\mu} = \Res_{w=0}
    \left(
      \re^{\frac{w^2}{4\pi\ri}+w\left(\bb(\delta+1/2+\lambda/A)
          +\bb^{-1}(\tilde{\delta}+1/2-\mu/A)\right)}
    \right)^A
    \left(\frac{\phi(\bb w,\delta_\bullet)
        \tilde{\phi}(\bb^{-1}w,\tilde{\delta}_\bullet)}
      {\bb(1-\re^{\bb^{-1}w})}\right)^B .
  \end{equation}
  \end{theorem}
  In particular, the right hand side of Equation~\eqref{thm1} is a
  bilinear combination of $q$ and $\tq$-series extend to the cut plane
  $\BC'$. A similar formula can be given when $\tau$ is in the lower
  half-plane, and what is more, the state integral satisfies the
  symmetry
  $\calI_{A,B,\lambda,\mu}(\bb) = \calI_{A,B,\lambda,\mu}(\bb^{-1})$
  which is a consequence of the symmetry
  $\Phi_{\bb^{-1}}(x)=\Phi_{\bb}(x)$ of the quantum dilogarithm.
  
  \begin{proof}
  We follow the derivation in \cite{GK:qseries} closely.
  The idea is to sum up residues at all singularities in the upper
  half-plane.  The factor $\Phi_{\bb}(x)$ has poles at
  \begin{equation}
    x_{m,n} = \ri\bb(m+1/2) + \ri\bb^{-1}(n+1/2),\quad m,n\in
    \BN \,.
  \end{equation}
  We notice that
  \begin{equation}
    \re^{2\pi(\lambda\bb - \mu\bb^{-1})(x+x_{m,n})} =
    \re^{w(\lambda\bb - \mu\bb^{-1})}
    q^{\lambda(m+1/2)}\tq^{\mu(n+1/2)}(-1)^{\lambda-\mu} 
  \end{equation}
  where we have made the change of variables $w = 2\pi x$ and used the
  notation
  \begin{equation}
    q = \re^{2\pi\ri \bb^2},\quad \tq = \re^{-2\pi\ri \bb^{-2}}.
  \end{equation}
  Now by modifying Eqn.(27) of~\cite{GK:qseries}, we find
  \begin{multline}
    \calI_{A,B,\lambda,\mu}(\bb) =
    (-1)^{\lambda-\mu} q^{\frac{\lambda}{2}}\tq^{\frac{\mu}{2}}
    \left(\frac{\tq}{q}\right)^{\frac{B-3A}{24}}
    \re^{\pi\ri\frac{B+2(A+1)}{4}} \times \\
    \sum_{m,n=0}^\infty \left(\Res_{w = 0}F_{A,B,m,n,\lambda,\mu}(w)\right)
    \frac{q^{\lambda m}t_m(q)^A}{(q;q)_m^B}
    \frac{\tq^{\mu n}\tilde{t}_n(\tq)^A}{(\tq^{-1};\tq^{-1})_n^B} \,,
  \end{multline}
  where
  \begin{equation}
    F_{A,B,m,n,\lambda,\mu}(w) =
    \left(e^{\frac{w^2}{4\pi\ri}+w\left(\bb(m+1/2
          +\lambda/A)+\bb^{-1}(n+1/2-\mu/A)\right)}
    \right)^A
    \left(\frac{\phi_m(\bb w)\tilde{\phi}_n(\bb^{-1}w)}
      {\bb(1-e^{\bb^{-1}w})}\right)^B \,.
  \end{equation}
  Using the operator formalism in \cite{GK:qseries}, this concludes
  the proof of Equation~\eqref{thm1}.
\end{proof}

  The reader may find in~\cite{GK:qseries} the expressions of the operators
  $\phi(w,\delta_\bullet)$, $\tilde{\phi}(w,\tilde{\delta}_\bullet)$.
  Note that $e_l(\tq)$ in the paper are simply $E_l^{(0)}(\tq)$.
  In addition,
  \begin{align}
    &F_{A,B,\lambda}(q,x) = \sum_{n=0}^\infty (-1)^{A n}
      \frac{q^{A\frac{n(n+1)}{2}+n\lambda}}{(q;q)_n^B} x^n,\\
    &\wt{F}_{A,B,\mu}(\tq,\tx) = \sum_{n=0}^\infty (-1)^{(B-A) n}
      \frac{\tq^{(B-A)\frac{n(n+1)}{2}+n\mu}}{(\tq;\tq)_n^B} \tx^n.
  \end{align}
  They can be related by
  \begin{equation}
    F_{A,B,\lambda}(q^{-1},x) = \wt{F}_{A,B,-\lambda}(q,x).
  \end{equation}

  \begin{example}
\label{ex.41s}
In this example we illustrate theorem~\ref{thm.state} with the
state integral $\calI_{1,2,\lambda,\mu}(\bb)$ associated to the $4_1$
knot~\cite{AK}. As we will see, this reproduces the $q$-series
$G^0_m(q)$ and $G^1_m(q)$ of~\eqref{gm} and~\eqref{Gm}.

Using
  \begin{equation}
    F_\lambda(q,x) = F_{1,2,\lambda}(q,x) = \wt{F}_{1,2,\lambda}(q,x)
    = \sum_{n=0}^\infty (-1)^n \frac{q^{\frac{n(n+1)}{2}+n\lambda}}{(q;q)_n^2}x^n.
  \end{equation}
  we find that 
  \begin{equation}
    P_{1,2,\lambda,\mu} =
    \frac{\bb}{2}(1+2\delta - 4\delta_1+2\lambda)
    -\frac{\bb^{-1}}{2}(1+2\tilde{\delta} - 4\tilde{\delta}_1 + 2\mu).
  \end{equation}
  It then follows that ($\tau = \bb^2$)
  \begin{equation}
  \label{desc-41}
    \calI_{1,2,\lambda,\mu}(\bb)
    =
    (-1)^{\lambda-\mu+1}\ri q^{\frac{\lambda}{2}}\tq^{\frac{\mu}{2}}
    \left(\frac{q}{\tq}\right)^{\frac{1}{24}}
    \left(\frac{\tau^{1/2}}{2}G^1_\lambda(q)G^0_\mu(\tq) -
      \frac{\tau^{-1/2}}{2}G^0_\lambda(q)G^1_\mu(\tq)\right),
  \end{equation}
  where we have used that
  \be
  \vev{F_{\lambda}(q,x)}=G^0_\lambda(q) , \qquad
  \vev{(1+2\lambda+2\delta-4\delta_1)F_\lambda(q,x)}=G^1_\lambda(q). 
    \end{equation}
\end{example}
In appendix~\ref{sec.A52}, we give the details for the state integral
$\calI_{2,3,\lambda,\mu}(\bb)$ of the $5_2$ knot.




\section{Computations}
\label{sec.compute}

\subsection{Resurgent analysis}
\label{sc:res}

In this section we briefly review some basic ingredients of resurgent
analysis. A detailed exposition may be found for example
in~\cite{abs,msauzin}. Given a Gevrey-1 series 
\begin{equation}
  \varphi(\tau)=\sum_{n \ge 0} a_n \tau^{n},\quad a_n = O(C^n n!),
\end{equation}
its Borel transform is defined by
\begin{equation}
  \widehat \varphi(\zeta)= \sum_{k \ge 0} {a_k \over k!} \zeta^k. 
\end{equation}
It is a holomorphic function in a neighborhood of the origin. In
favorable cases, this function can be extended to the complex
$\zeta$-plane (also called Borel plane), but it will have
singularities. Assuming that the analytically continued function does
not grow too fast at infinity, the Borel resummation of
$\varphi(\tau)$ is defined as the Laplace transform
\begin{equation}
  s(\varphi)(\tau) = \int_0^\infty \re^{-\zeta}
  \widehat \varphi(\zeta \tau) \rd \zeta. 
\end{equation}
This has discontinuities at Stokes rays in the $\tau$ plane, whenever
$ {\rm arg}(\tau)= {\rm arg}(\zeta_s)$, where $\zeta_s$ is a
singularity of $\widehat \varphi(\zeta)$. We define the lateral Borel
resummations for $\tau$ near a Stokes ray by
\begin{equation}
  s_\pm (\varphi)(\tau)= \int_0^{\re^{\pm \ri \epsilon} \infty}
  \re^{-\zeta} \widehat \varphi(\zeta \tau) \rd \zeta. 
\end{equation}

In the context of the theory of resurgence, we are usually given a
collection of transseries $\Phi_\omega(\tau)$, where $\omega$ belongs
to an indexing set. These transseries have the form
\begin{equation}
  \Phi_\omega(\tau)=\re^{-V_\omega /\tau} \varphi_\omega(\tau),
  \qquad \varphi_\omega(\tau) \in \BC[[\tau]], 
\end{equation}
where $V_\omega$ is the ``action" associated to the sector $\omega$.
The Borel resummation of the trans-series $\Phi_\omega (\tau)$ is
defined by
\begin{equation}
  s(\Phi_\omega)(\tau) =\re^{-V_\omega /\tau} s(\varphi_\omega)(\tau)
\end{equation}
(with suitable care for the constant term of $\varphi_\omega(\tau)$).
To measure the discontinuity of Borel resummations across a Stokes
ray, one introduces the Stokes automorphism $\mathfrak{S}$ as
\begin{equation}
  s_+= s_- \mathfrak{S}. 
\end{equation}
In our case, the singularities of $\widehat \varphi(\tau)$ are
logarithmic branch points (i.e. we are dealing with so-called simple
resurgent functions).  In that case, the Stokes automorphism can be
expressed as a (possibly infinite) linear combination of transseries,
\begin{equation}
  \mathfrak{S} (\Phi_\omega)= \Phi_{\omega}+ \sum_{\omega'}
  \mathsf{S}_{\omega \omega'} \Phi_{\omega'}. 
\end{equation}
The coefficients $\mathsf{S}_{\omega \omega'}$ are the Stokes
constants (note that with this convention, their signs are opposite to
e.g. the ones in \cite{abs}.) The singularities of
$\widehat \varphi_\omega(\tau)$ occur at the points $\omega'-\omega$
for which $\mathsf{S}_{\omega \omega'} \not=0$.

In the case that we consider in this paper, the transseries are labeled 
by the critical point $\sigma$ and the multicovering $\mu \in \IZ$, i.e.
$\omega=(\s, \mu)$. If there is a singularity in the Borel plane of
$\Phi_{\s,\mu}$ located at
\begin{equation}
  \iota_{\s,\s'}^{(\mu,\lambda)} = \text{CS}(\s;\mu) - \text{CS}(\s';\lambda),
\end{equation}
representing another transseries $\Phi_{\s',\lambda}$, then the Borel
resummation 
$s(\Phi_{\s,\mu})(\tau)$ 
is discontinuous across the Stokes ray $\rho_{\theta}$ with
$\theta = \arg \iota_{\s,\s'}^{(\mu,\lambda)}$, and the associated
Stokes automorphism reads
\begin{equation}
  s_{+}(\Phi_{\s,\mu})(\tau) =
  s_{-}(\Phi_{\s,\mu})(\tau)
  + \ms{S}_{\s,\s'}^{(\mu,\lambda)}
  s_{-}(\Phi_{\s',\lambda})(\tau),
  \label{eq:stokes-auto}
\end{equation}
where 
$\ms{S}_{\s,\s'}^{(\mu,\lambda)}$ is the Stokes constant.  Equation
\eqref{Phiqt} implies that
$\ms{S}_{\s,\s'}^{(\mu,\lambda)} = \ms{S}_{\s,\s'}^{(\lambda-\mu)}$
depends on $\s,\s'$ and the difference $\lambda-\mu$, an arbitrary
integer number.  It follows the equation of Stokes automorphism
\eqref{eq:stokes-auto} can be written as
\begin{equation}
  s_{+}(\Phi)(\tau) = \mf{S}_{\theta}(\tq)
  s_{-}(\Phi)(\tau)
  \label{eq:Sauto-mat}
\end{equation}
with $\Phi(\tau)=(\Phi_\s(\tau))_\s$ the vector of asymptotic series,
and the Stokes automorphism matrix
\begin{equation}
  \mf{S}_{\theta}(\tq) = I + \ms{S}_{\s,\s'}^{(k)}\tq^k E_{\s,\s'}
  \label{eq:Sth}
\end{equation}
where $E_{\s,\s'}$ is the elementary matrix with $(\s,\s')$-entry 1
and all other entries zero.  Furthermore, all the Stokes constants are
encoded in the two Stokes matrices
\begin{equation}
  \ms{S}^+(\tq) = \mf{S}_{-\epsilon \to \pi-\epsilon}(\tq),\quad
  \ms{S}^-(\tq^{-1}) = \mf{S}_{\pi-\epsilon\to 2\pi-\epsilon}(\tq)
  \label{eq:Spn}
\end{equation}
(for $\epsilon>0$ and sufficiently small) where
$\mf{S}_{\theta^-\to\theta^+}$ is the global Stokes automorphism matrix
defined for two non-Stokes rays whose arguments satisfy
$0\leq \theta^+-\theta^- \leq \pi$ by
\begin{equation}
  \mf{S}_{\theta^-\to \theta^+}(\tq) =
  \prod_{\theta^-<\theta<\theta^+}^{\longleftarrow}
  \mf{S}_{\theta}(\tq)
  \label{eq:Sfac}
\end{equation}
where the ordered product is taken over the Stokes rays in the cone
generated by $\rho_{\theta^-}$ and $\rho_{\theta^+}$.  This
factorization is well-known in the classical literature on the WKB
method (see for example \cite{voros} where it is called the ``radar
method''), and we will discuss it in more detail including its
uniqueness in \cite{GGM-peacock}.
%
%
Note the Stokes automorphisms are now represented by two
finite-dimensional matrices $\ms{S}^+(\tq)$ and $\ms{S}^-(\tq^{-1})$ in
which the entries have been promoted to $\tq$ and $\tq^{-1}$-series
respectively. This reorganization of the transseries is reminiscent of
what was done in~\cite{costin-costin}.

To numerically compute the integer coefficients of the above
$\tq$-series, we need a high precision numerical computation of the
Laplace integrals.  And here lies the issue.  In practice, only a few
hundred coefficients of the series $\varphi(\tau)$ can be
obtained. For instance, for the $4_1$ knot, the stationary phase of
the state integral allows one to compute 300 coefficients of
$\varphi_{\sigma_{1,2}}(\tau)$, and for the $5_2$ knot about 200
coefficients of $\varphi_{\sigma_{1,2,3}}(\tau)$ can be
obtained. Alternatively, a numerical computation of the Kashaev
invariant together with numerical extrapolation gives about 100
terms. Given such a truncated series, one can use Pad\'e approximants
to analytically continue the Borel transform to the complex plane, and
then calculate the Borel resummation numerically. The Pad\'e
approximant can be also used to determine numerically the
singularities in the Borel plane. Precision can be improved by using a
conformal mapping, see \cite{caliceti} for a summary of numerical
techniques.

\subsection{The $4_1$ knot}
\label{sub.asy41}

\begin{figure}[!ht]
\leavevmode
\begin{center}
\includegraphics[height=7cm]{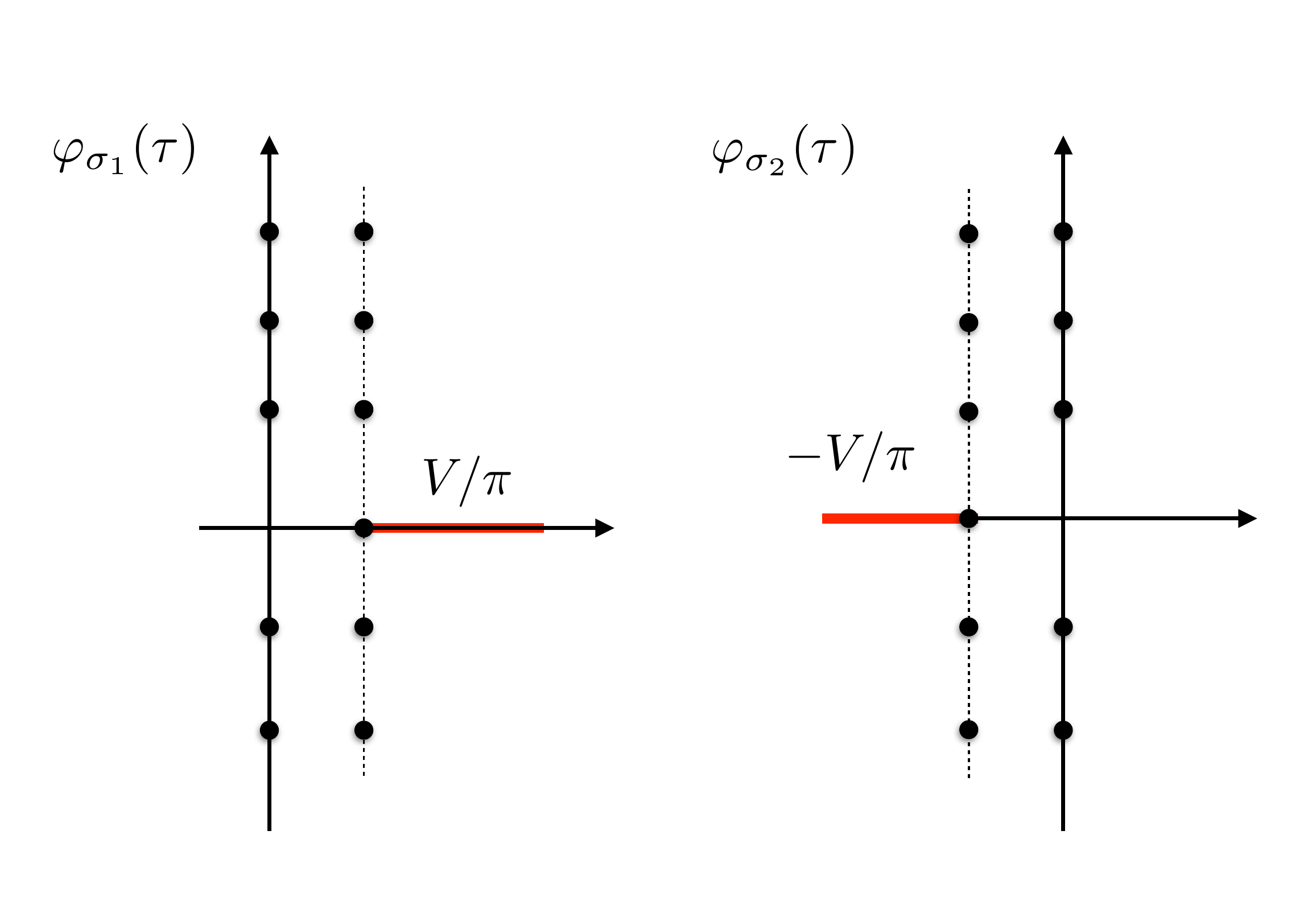}
\end{center}
\caption{The singularities in the Borel plane for the series
  $\varphi_{\sigma_{1,2}} (\tau)$.}
\label{bpf8-fig}
\end{figure} 

\begin{figure}[!ht]
\leavevmode
\begin{center}
\includegraphics[height=9cm]{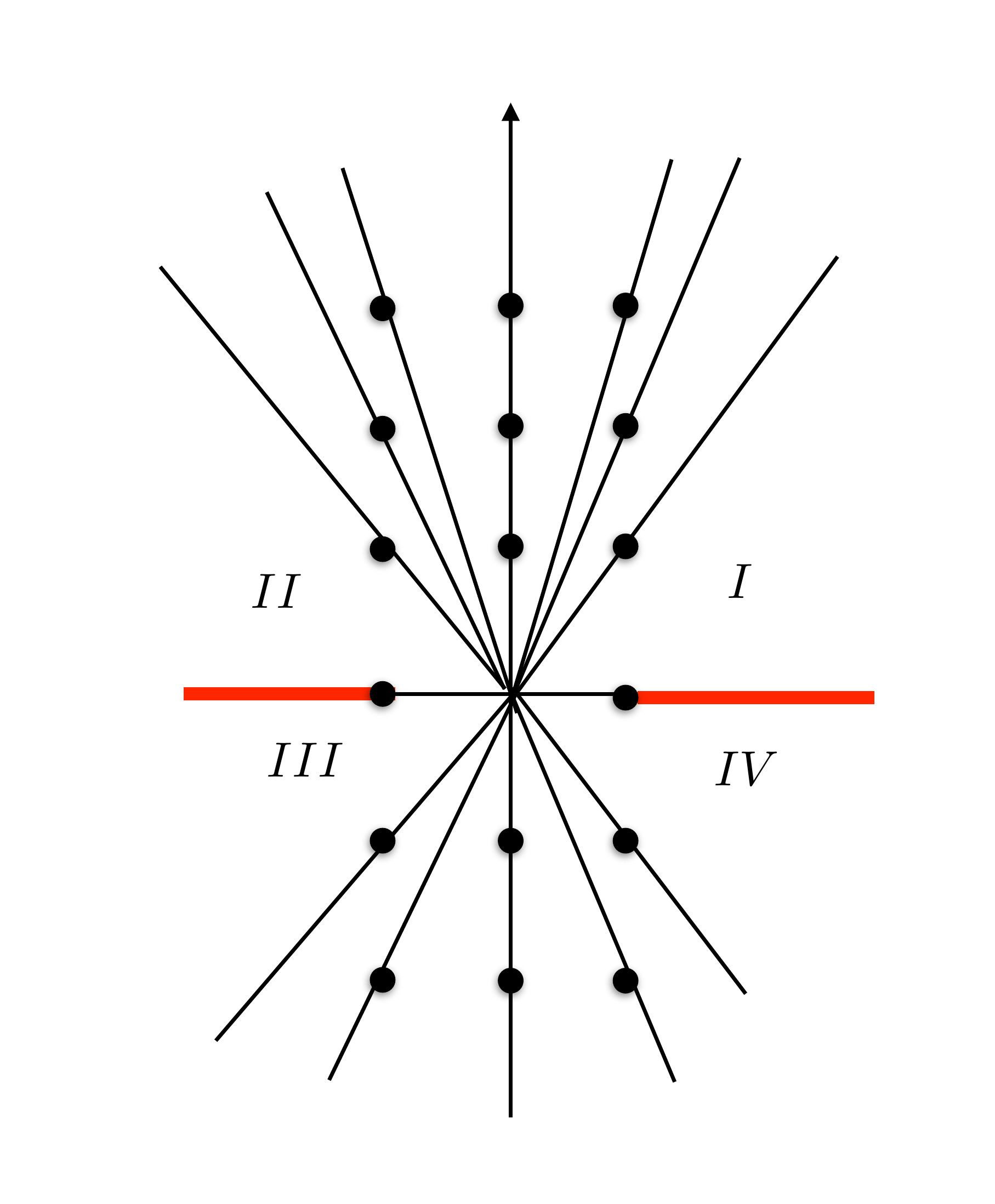}
\end{center}
\caption{Stokes rays in the $\tau$-plane for $\Phi(\tau)$.}
\label{ray-fig}
\end{figure}

The structure of singularities in the Borel plane for the formal power
series $\varphi_{1,2}(\tau)$ of the $4_1$ knot is shown
in~\figref{bpf8-fig}.  Points in each vertical line are $2\pi\ri$
apart, and the two points in the real axis correspond to
\begin{equation}
\iota_{\pm} = \pm \left( {V(\s_1)\over 2 \pi} -{V(\s_2)\over 2 \pi} \right)=
\pm {V \over \pi}, 
\end{equation}
where $V$ is defined in \eqref{volume-41}. Since each singularity in
the Borel plane leads to a discontinuity in the Borel resummation, one
has the structure of Stokes rays shown in \figref{ray-fig} (where we
took into account both series). Note that there is an infinite dense
set of rays accumulating towards the imaginary axis.

We already hinted at the end of Section~\ref{sub.gG41} that the
asymptotic expansion~\eqref{gGtau} can be upgraded to an {\it exact}
expression.  To do this, one has to upgrade the optimal truncation of
$\Phi(\tau)$ to its Borel resummation. Simultaneously we have to
promote the matrix of constants appearing in ~\eqref{gGtau}, to a
matrix whose entries are power series in $\tq$ with integer
coefficients:
\begin{equation}
\label{MR}
\begin{pmatrix} \frac{1}{\sqrt{\tau}} g^0_0(\tau) \\
  \sqrt{\tau} g^1_0(\tau)
\end{pmatrix} = M_R (\tq) \, s_R(\Phi)(\tau). 
\end{equation}
The index $R$ labels a sector in the $\tau$-plane, since due to
presence of Stokes rays, both the matrix $M_R(\tq) $ and the Borel
resummed vector $\Phi$ depend on the sector of the $\tau$-plane.  In
view of the structure of the Stokes rays, convenient sectors to
perform the analysis are the angular wedges (i.e., pointed open cones
in the complex plane) denoted by $I$, $II$, $III$ and $IV$
in~\figref{ray-fig}.

It is a challenge to numerically compute the matrix $M_R(\tq)$ given
only a few hundred terms of $\Phi(\tau)$, since the volume of $4_1$
(about $2.02\dots$) is so much smaller than the instanton corrections
(appearing at $4\pi^2 = 39.47\dots$). This can be done however, and
with 300 terms of $\Phi(\tau)$ it is possible to compute the first
twelve terms in the series appearing in $M_R(\tq)$.  One finds for
example, in region $I$,
\begin{equation}
\label{M0val}
M_I(q) = 
\begin{pmatrix} 
  1 - q - 2 q^2 - 2 q^3 - 2 q^4 
  &  -1 + 2 q + 3 q^2 + 2 q^3 + q^4 
\\
1 - 7 q - 14 q^2 - 8 q^3 - 2 q^4  
& 
1 + 10 q + 15 q^2 - 2 q^3 - 19 q^4 
\end{pmatrix} +O(q^5).  
\end{equation}
Our Conjecture \ref{conj.1} suggests that these $q$-series can be
expressed in terms of solutions to the linear $q$-difference equation
\eqref{41qdiff}.  Indeed, one has, at this order,
\begin{equation}
M_I(q)=
\begin{pmatrix} G_0^0 (q) & -G_0^0(q) - G_{-1}^0(q) \\
  G_0^1 (q) &  -G_0^1(q) - G_{-1}^1(q)
\end{pmatrix}
= W_{-1}(q)^T \,
    \begin{pmatrix}
      0 & -1 \\
      1 & -1
    \end{pmatrix}
\,.
\end{equation}
%
We conjecture that this is in fact the exact expression
for this matrix.

An important consequence of the relation~\eqref{MR} is that, by
inverting it, one can express the Borel resummations of
$\Phi_{\s_{j}}$ for $j=1,2$ in a given sector, in terms of the
descendants of the state integral introduced in~\eqref{sub.state},
which are holomorphic functions of $\tau$ on $\BC'$. Indeed, let us
consider the ``reduced" descendant
\begin{equation}
 \widetilde \calI_{1,2,\lambda,\mu}(\bb)
    =   \frac{\tau^{1/2}}{2}G^1_\lambda(q)G^0_\mu(\tq) -
      \frac{\tau^{-1/2}}{2}G^0_\lambda(q)G^1_\mu(\tq). 
\end{equation}
This differs from the descendant (\ref{desc-41}) in a manifestly
holomorphic factor, so it is holomorphic on $\BC'$. Then, one finds,
in region $I$,
\begin{equation}
s_I(\Phi_{\s_1})(\tau)=  \widetilde \calI_{1,2,0,0}(\tau)
+  \widetilde \calI_{1,2,0,-1,}(\tau), \qquad 
s_I(\Phi_{\s_2})(\tau)=  \widetilde \calI_{1,2,0,0}(\tau). 
\end{equation}

This procedure can be done in the other sectors appearing
in~\figref{ray-fig}: one calculates $M_R(q)$, express it in terms of
fundamental solutions, and represent the Borel resummation $s_R(\Phi)$
in terms of holomorphic functions on $\BC'$. By comparing the
different expressions for the Borel resummations in different sectors,
one deduces the Stokes automorphisms relating them, and from a
composition of the Stokes automorphisms one deduces the promised
$\mathsf{S}$ matrices.
 
The results for $M_R(q)$ are the following:
\begin{equation}
\begin{aligned}
   M_{II}(q)&=\begin{pmatrix} G_0^0(q) + G_{-1}^0(q)& -G_0^0(q)  \\
     -G_0^1(q) - G_{-1}^1(q) & G_0^1 (q)    \end{pmatrix} =
\begin{pmatrix}
    1 & 0 \\
    0 & -1
  \end{pmatrix} \, W_{-1}(q)^T \,
        \begin{pmatrix}
          1 & 0 \\
          1 &-1
        \end{pmatrix}
   , \\
   M_{III}(q)&=\begin{pmatrix} G_0^0(q^{-1}) + G_{-1}^0(q^{-1})& G_0^0(q^{-1})
     \\ G_0^1(q^{-1}) +G_{-1}^1(q^{-1}) & G_0^1 (q^{-1})    \end{pmatrix}
   =
    W_{-1}(q^{-1})^T \,
    \begin{pmatrix}
      1 & 0 \\
      1 & 1 \end{pmatrix},\\
 M_{IV}(q)&=\begin{pmatrix} G_0^0(q^{-1}) &  G_0^0(q^{-1}) + G_{-1}^0(q^{-1}) 
   \\ -G_0^1 (q^{-1}) & -G_0^1(q^{-1}) -G_{-1}^1(q^{-1})      \end{pmatrix}
 =\begin{pmatrix}
    1 & 0 \\
    0 & -1
  \end{pmatrix} \,
        W_{-1}(q^{-1})^T \,
        \begin{pmatrix}
          0 & 1 \\
          1 & 1
        \end{pmatrix} \,. 
\end{aligned}
\end{equation}
From these values one deduces the Stokes automorphisms
%
\begin{align}
  s_{II}(\Phi)&= \mathfrak{S}_{I \rightarrow II}(\tq)s_{I}(\Phi),
  & s_{IV}(\Phi)&= \mathfrak{S}_{III \rightarrow IV}(\tq^{-1})s_{III}(\Phi)\nn
s_{I}(\Phi) &= \mf{S}_{IV\mapsto I} s_{IV}(\Phi), &
s_{III}(\Phi) &= \mf{S}_{II\mapsto III} s_{III}(\Phi),
\end{align}
where
\begin{subequations}
  \begin{align}
  \label{Spm3}
    \mathfrak{S}_{I \rightarrow II}(q)
    &= M_{II}(q)^{-1} M_I(q)
    & \mathfrak{S}_{III \rightarrow IV}(q)
    &= M_{IV}(q^{-1})^{-1}
  M_{III}(q^{-1}) \\
  \label{Spm4}
    \mathfrak{S}_{IV\rightarrow I}
    & = M_I(q)^{-1}M_{IV}(q)
    & \mathfrak{S}_{II\rightarrow III}
    &= M_{III}(q)^{-1}M_{II}(q) \,.
\end{align}
\end{subequations}
and
\begin{equation}
  \label{Spm1}
  \ms{S}^+(q)= 
  \mathfrak{S}_{I \rightarrow II}(q)
  \mathfrak{S}_{IV\rightarrow I}, \qquad
  \ms{S}^-(q)= 
  \mathfrak{S}_{III \rightarrow IV}(q)
  \mathfrak{S}_{II\rightarrow III}\,.
\end{equation}
Substituting the conjectured values for $M_R$
and using symmetry and the orthogonality relations
~\eqref{W41inv} and~\eqref{WWT41b} 
one obtains \eqref{S41p}, \eqref{S41m}.
In the $q\mapsto 0$ limit the Stokes matrices read
\begin{equation}
  \ms{S}^+(0) =
  \begin{pmatrix}
    1&3\\0&1
  \end{pmatrix}\,,\quad
  \ms{S}^-(0) =
  \begin{pmatrix}
    1&0\\-3&1
  \end{pmatrix}.
\end{equation}
The off--diagonal entries $\pm 3$ are Stokes constants associated to
the singularities $\iota_{\pm}$ on the positive and negative real axis
respectively.  They agree with the matrix of integers obtained in
\cite{gh-res},\cite{GZ:kashaev}. Note that the Stokes matrices
$\ms{S}^{\pm}(q)$ can also be factorized according to \eqref{eq:Spn}
in order to extract all the other Stokes constants.  This will be
studied in \cite{GGM-peacock}. Let us finally note that 
\begin{equation}
\mathsf{S}^+_{\sigma_1 \sigma_1}(q)= G_0^0(q) G_0^1  (q)
=1 -8 q -9 q^2 +18 q^3 + 46 q^4+ O(q^5)
= {\rm Ind}^{\rm rot}_{4_1}(q), 
\end{equation}
in agreement with Conjecture~\ref{conj.2} (the fact that
$ G_0^0(q) G_0^1 (q)$ equals the rotated index was pointed out
in~\cite{GZ:qseries}).

\subsection{The $5_2$ knot}
\label{sub.asy52}

\begin{figure}[!ht]
\leavevmode
\begin{center}
  \includegraphics[height=8cm]{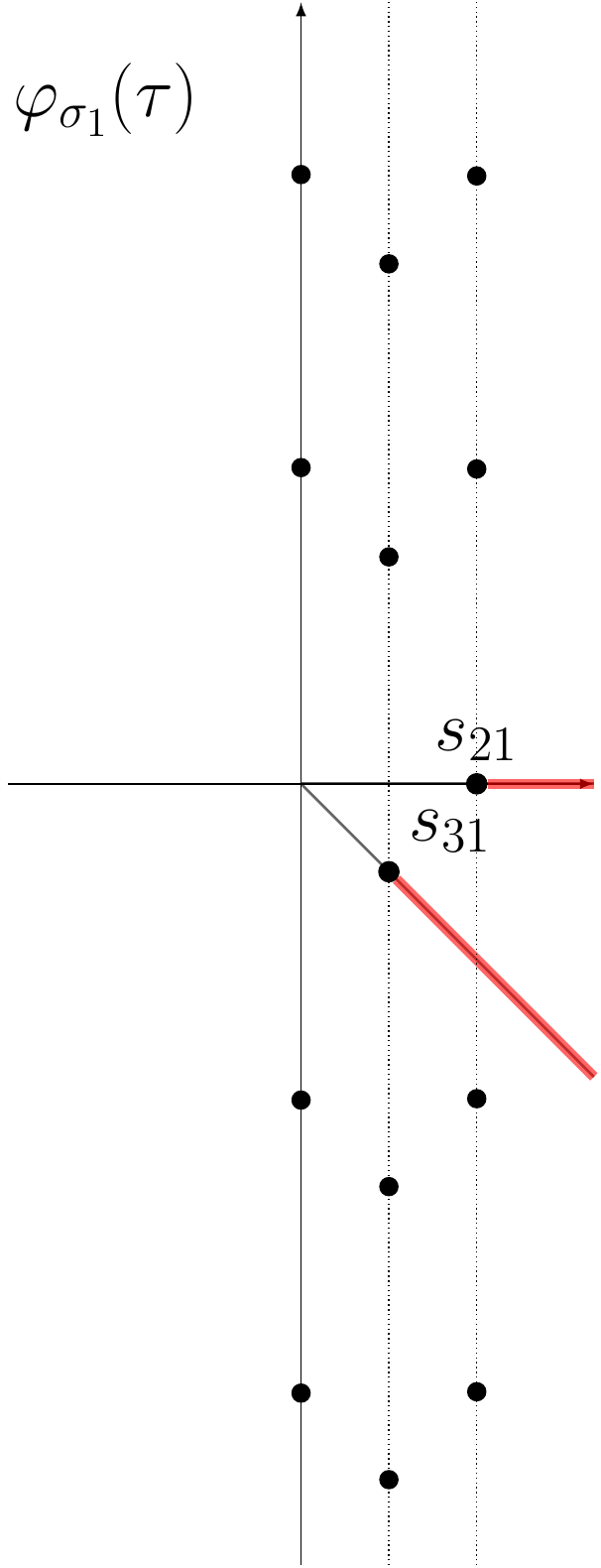}
  \hspace{5ex}
  \includegraphics[height=8cm]{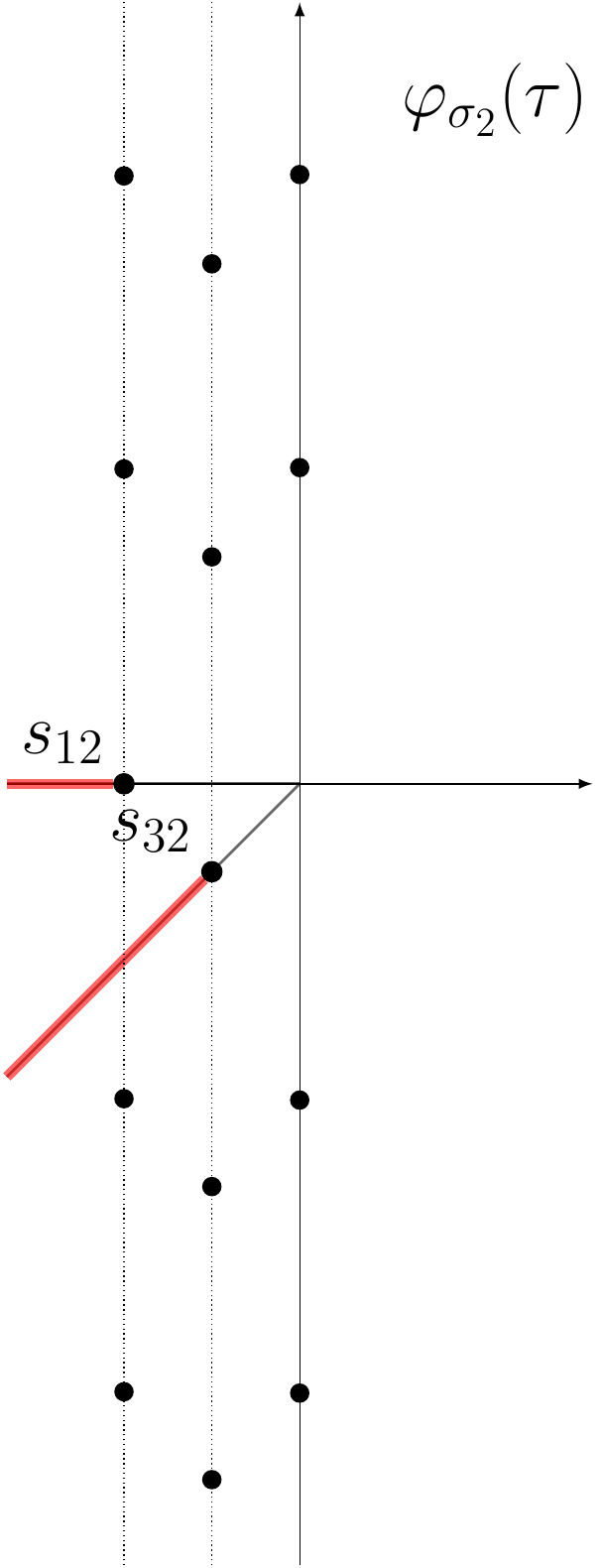}
  \hspace{5ex}
  \includegraphics[height=8cm]{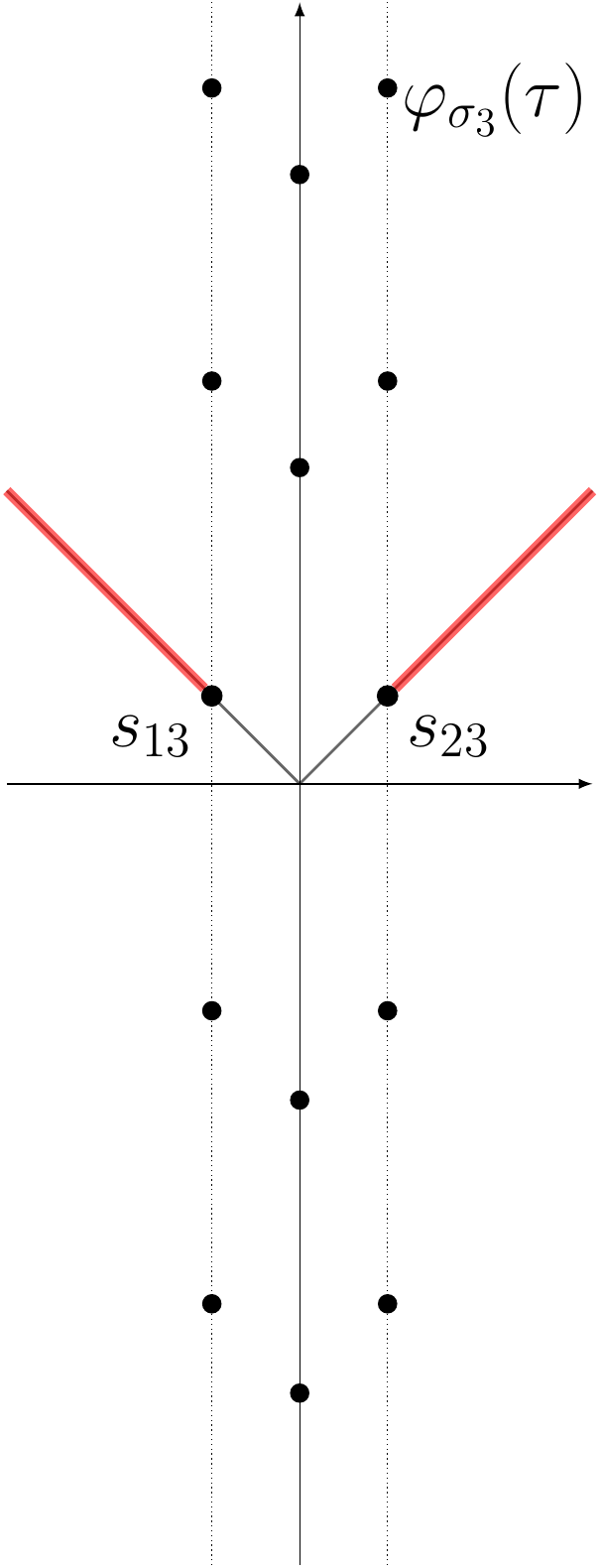}
\end{center}
\caption{The singularities in the Borel plane for the series
  $\varphi_{\sigma_{j}} (\tau)$ of the $5_2$ knot for $j=1,2,3$.}
\label{bp3t-fig}
\end{figure}

\begin{figure}[!ht]
\leavevmode
\begin{center}
\includegraphics[width=0.3\linewidth]{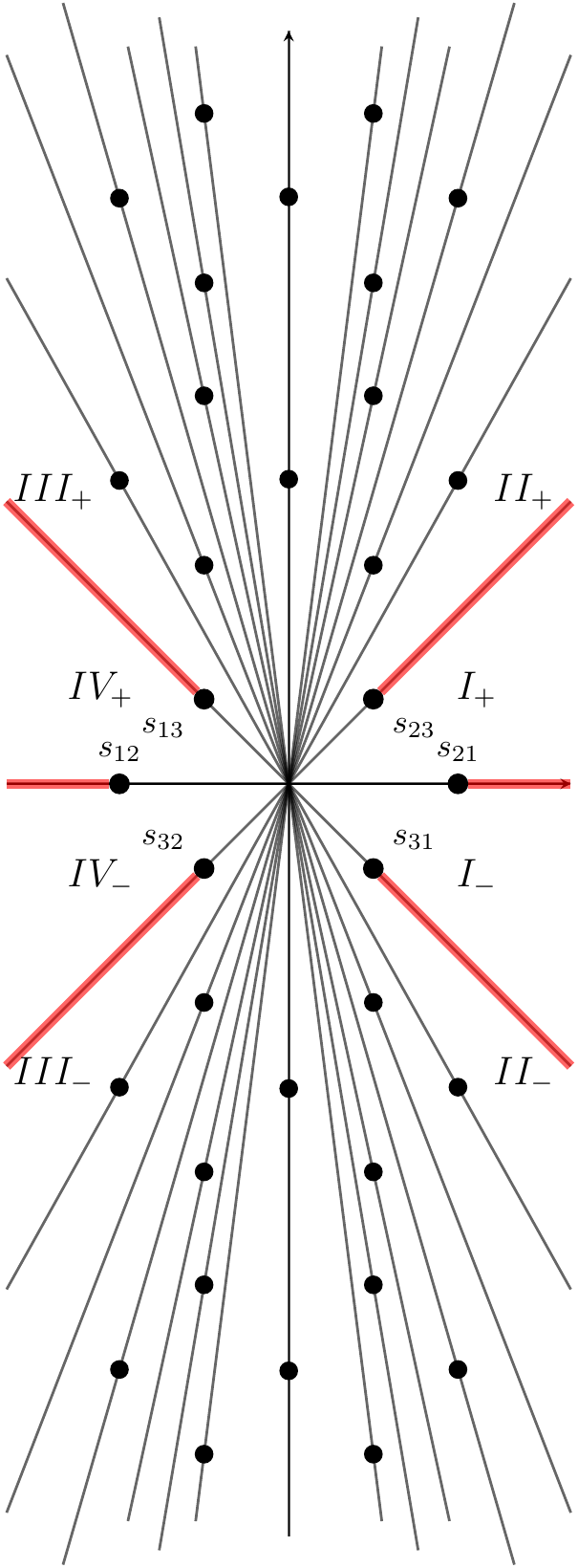}
\end{center}
\caption{Stokes rays in the $\tau$-plane for $\Phi_{\sigma_{j}}(\tau)$
  of the $5_2$ knot for $j=1,2,3$.  Note that the points
  $s_{23},s_{31}$ happen to have the same real part, and so do
  $s_{13},s_{32}$.  The dots are not shown in scale for aesthetic
  purpose.}
\label{3tray-fig}
\end{figure}

The structure of singularities in the Borel plane for the formal power
series $\varphi_{\sigma_j}(\tau)$ ($j=1,2,3$) of the $5_2$ knot is
shown in Fig.~\ref{bp3t-fig}.  Points in each vertical line are
$2\pi\ri$ are apart, while the six points $\iota_{ij}$ surrounding the
origin are given by
\begin{equation}
  \iota_{ij} = \frac{V(\sigma_i)}{2\pi} - \frac{V(\sigma_j)}{2\pi},\qquad
  1 \leq i\neq j \leq 3 \,,
\end{equation}
where $V(\sigma_i)$ are given in \eqref{eq:V_52}.  The structure of
Stokes rays is shown in Fig.~\ref{3tray-fig}.  There is also an
infinite dense set of rays accumulating towards the imaginary axis.

The $q$-series $H^{\pm}_{k,0}$ for $(k=0,1,2)$, which are analogues of
$G^0_0,G^1_0$ of the $4_1$ knot, have similarly interesting radial
asymptotics.
We use small letters for the corresponding functions, i.e.
\begin{equation}
  h_k(\tau) =
  \begin{cases}
    H^+_{k,0}(\re^{2\pi\ri\tau}),\quad & \Im(\tau) > 0\\
    H^-_{k,0}(\re^{-2\pi\ri \tau}),\quad & \Im(\tau) < 0
  \end{cases},\quad k=0,1,2 \,.
  \label{eq:hk}
\end{equation}
Then the exact expression of the radial asymptotics reads
\begin{equation}
  \re^{3\pi\ri/4}
  \left(
    \begin{array}{c}
      \tau^{-1} h_{0}(\tau)\\
      h_{1}(\tau)\\
      \tau h_{2}(\tau)
    \end{array}\right) = M_R(\tq) s_R(\Phi)(\tau),
\end{equation}
where the index $R$ labels a sector in the $\tau$-plane. The entries
of the matrix $M_R(\tq)$ are power series in $\tq$ in the upper half
plane, and power series in $1/\tq$ in the lower half-plane.

With more than 200 terms of $\Phi(\tau)$, we are able to compute first
few terms in $M_R(\tq)$.  For instance, in region $I_+$ we are able to
compute first six terms in each entry of $M_{I_+}(\tq)$.  We display
some of the results here
\begin{equation}
  M_{I_+}(q) =
  \begin{pmatrix}
    -1 - q^2  &  2 + 3 q^2 & 1 + q + 3 q^2 \\
    -1 + 3q + 3q^2 & 1 - 6q - 3 q^2 &-q \\
    -\frac{5}{6} + 5 q -\frac{53}{6}q^2& -\frac{4}{3} - 4q +
    \frac{77}{2}q^2 &-\frac{1}{6} + \frac{29}{6}q
    +\frac{55}{2}q^2
  \end{pmatrix} +\cO(q^3).
\end{equation}
Following Conjecture \ref{conj.1}, we can express it in terms of
solutions to the linear $q$-difference equation
\eqref{52qdiff}
\begin{equation}
  M_{I_+}(q) = W_{-1}(q)^T
  \begin{pmatrix}
    0&0&1\\
    -1&3&0\\
    0&-1&0
  \end{pmatrix},
\end{equation}
where the Wronskian is defined in \eqref{52fund1}. Note that this
expression is exact. By inverting the matrix $M_{I_+}(q)$, we can
express the Borel resummation of $\Phi_{\sigma_{j}}$ for $j=1,2,3$ in
the sector $I_+$ in terms of the descendants of the state integal
introduced in Equation~\eqref{eq.statelm}.  Let us again introduce the
``reduced'' descendant
\begin{equation}
  \wt{\mc{I}}_{\lambda,\mu}(\tau) =
  \tau H_{2,\lambda}^{+}(q)H_{0,\mu}^{-}(\tq)
  -2 H_{1,\lambda}^{+}(q)H_{1,\mu}^{-}(\tq)
  +\tau^{-1} H_{0,\lambda}^{+}(q)H_{2,\mu}^{-}(\tq),
\end{equation}
which differs from the descendant \eqref{eq:I23lm} in a manifestly
holomorphic factor. We find in region $I_+$
\begin{equation}
  s_{I_+}(\Phi)(\tau)
  =
  \frac{1}{2}\re^{3\pi\ri/4}
    \begin{pmatrix}
      -1 & -1 & 0 \\ 0 & -1 & 0\\0 & 0 & 1
    \end{pmatrix}
     \left(
      \begin{array}{c}
        \wt{\mc{I}}_{0,-1}\\
        \wt{\mc{I}}_{0,0}\\
        \wt{\mc{I}}_{0,1}\\
      \end{array}\right)(\tau)\,.
\end{equation}
In other regions, the results for $M_R(q)$ are as follows: 
\begin{itemize}
\item In the upper half-plane
  \begin{align}
    M_{II_+}(q) =
    &W_{-1}(q)^T
      \begin{pmatrix}
        0&-3&1\\
        -1&3&0\\
        0&-1&0
      \end{pmatrix},\\
    M_{III_+}(q) = &\begin{pmatrix}
      1&0&0\\
      0&-1&0\\
      0&0&1
    \end{pmatrix}
           W_{-1}(q)^T
  \begin{pmatrix}
    -3&0&1\\
    3&-1&0\\
    -1&0&0
  \end{pmatrix}, \\
      M_{IV_+}(q) =
  &\begin{pmatrix}
      1&0&0\\
      0&-1&0\\
      0&0&1
  \end{pmatrix}
  W_{-1}(q)^T
  \begin{pmatrix}
    0&0&1\\
    3&-1&0\\
    -1&0&0
  \end{pmatrix},\qquad |q|<1.
  \end{align}
\item In the lower half-plane:
 \begin{align}
  M_{I_-}(q) = &\begin{pmatrix}
      1&0&0\\
      0&-1&0\\
      0&0&1
  \end{pmatrix}W_{-1}(q)^T
  \begin{pmatrix}
      0&0&1\\
      -1&-1&0\\
      0&-1&0
  \end{pmatrix}, \\
  M_{II_-}(q) = &\begin{pmatrix}
      1&0&0\\
      0&-1&0\\
      0&0&1
  \end{pmatrix}W_{-1}(q)^T
  \begin{pmatrix}
    0&0&1\\
    -1&-1&-3\\
    0&-1&0
  \end{pmatrix},\\
  M_{III_-}(q) =
  &W_{-1}(q)^T
  \begin{pmatrix}
    0&0&1\\
    -1&-1&-3\\
    -1&0&0
  \end{pmatrix},\\
  M_{IV_-}(q) =
  &W_{-1}(q)^T
  \begin{pmatrix}
    0&0&1\\
    -1&-1&0\\
    -1&0&0
  \end{pmatrix},\qquad |q|>1 .
 \end{align}
 
\end{itemize}
From these values one can deduce the Stokes automorphism, in the
anticlockwise direction
\begin{gather}
  s_{III_+}(\Phi) = \mf{S}_{II_+\to III_+}(\tq) s_{II_+}(\Phi),\quad
  s_{II_-}(\Phi) = \mf{S}_{III_-\to II_-}(\tq^{-1}) s_{III_-}(\Phi),\nn
  s_{II_+}(\Phi) = \mf{S}_{I_+\to II_+} s_{I_+}(\Phi),\quad
  s_{I_-}(\Phi) = \mf{S}_{II_-\to I_-} s_{II_-}(\Phi),\nn
  s_{IV_+}(\Phi) = \mf{S}_{III_+\to IV_+} s_{III_+}(\Phi),\quad
  s_{III_-}(\Phi) = \mf{S}_{IV_-\to III_-} s_{IV_-}(\Phi),\nn
  s_{I_+}(\Phi) = \mf{S}_{I_-\to I_+} s_{I_-}(\Phi),\quad
  s_{IV_-}(\Phi) = \mf{S}_{IV_+\to IV_-} s_{IV_+}(\Phi),
\end{gather}
where
\begin{align}
  \mf{S}_{II_+\to III_+}(q) =
  &\frac{1}{2}
      \begin{pmatrix}
        0&-1&0\\
        0&-1&-1\\
        1&-3&0
      \end{pmatrix}\cdot W_{-1}(q^{-1})\cdot
               \begin{pmatrix}
                 0&0&1\\
                 0&-2&0\\
                 1&0&0
               \end{pmatrix}\cdot W_{-1}(q)\cdot
               \begin{pmatrix}
                 0&-3&1\\
                 -1&3&0\\
                 0&-1&0
               \end{pmatrix},\\
  \mf{S}^{-1}_{III_-\to II_-}(q) =
  &\frac{1}{2}
      \begin{pmatrix}
        0&-1&0\\
        -3&3&-1\\
        1&0&0
      \end{pmatrix}
              \cdot W_{-1}(q) \cdot
              \begin{pmatrix}
                 0&0&1\\
                 0&-2&0\\
                 1&0&0
               \end{pmatrix}\cdot W_{-1}(q^{-1})^T
                      \cdot
              \begin{pmatrix}
                0&0&1\\
                -1&-1&-3\\
                0&-1&0
              \end{pmatrix},
\end{align}
and $\mf{S}_{I_-\to I_+}$,$\mf{S}_{I_+\to
  II_+}$,$\mf{S}_{III_+\to
  IV_+}$,$\mf{S}_{IV_+\to
  IV_-}$,$\mf{S}_{IV_-\to III_-}$,$\mf{S}_{II_-\to I_-}$ are given in
\eqref{eq:S0pu052},\eqref{eq:S0nu052}. The matrices
$\ms{S}^{\pm}(q)$ are simply
\begin{align}
  \ms{S}^+(q) = &\mf{S}_{III_+\to IV_+}
                  \mf{S}_{II_+\to III_+}(q)\mf{S}_{I_+\to II_+}\mf{S}_{I_-\to I_+},\\
  \ms{S}^-(q) = &\mf{S}_{II_-\to I_-}\mf{S}_{III_-\to
                  II_-}(q)\mf{S}_{IV_-\to III_-} \mf{S}_{IV_+\to IV_-},
\end{align}
and we obtain \eqref{S52p},\eqref{S52m}.

In the $q\mapsto 0$ limit, the Stokes matrices factorize
\begin{subequations}
\begin{align}
  \ms{S}^+(0) =
  &\mf{S}_{\s_3,\s_1}\mf{S}_{\s_3,\s_2}\mf{S}_{\s_1,\s_2}
    =
     \begin{pmatrix}
       1&0&0\\0&1&0\\-3&0&1
     \end{pmatrix}
     \begin{pmatrix}
       1&0&0\\0&1&0\\0&3&1
     \end{pmatrix}
     \begin{pmatrix}
      1&4&0\\0&1&0\\0&0&1
    \end{pmatrix},  \label{eq:S0pu052}\\
  \ms{S}^-(0) =
  &\mf{S}_{\s_1,\s_3}\mf{S}_{\s_2,\s_3}\mf{S}_{\s_2,\s_1}
    =
     \begin{pmatrix}
       1&0&3\\0&1&0\\0&0&1
     \end{pmatrix}
     \begin{pmatrix}
       1&0&0\\0&1&-3\\0&0&1
     \end{pmatrix}
     \begin{pmatrix}
      1&0&0\\-4&1&0\\0&0&1
    \end{pmatrix}. \label{eq:S0nu052}
\end{align}
\end{subequations}
The non-vanishing off-diagonal entry of $\mf{S}_{\s_i,\s_j}$ is the
Stokes constant associated to the Borel singularity $\iota_{i,j}$.
Assembling these Stokes constants in a matrix we obtain
\begin{equation}
  \begin{pmatrix}
    0&4&3\\
    -4&0&-3\\
    -3&3&0
  \end{pmatrix}
\end{equation}
which is what was found numerically in \cite[Sec.3.3]{GZ:kashaev}.
Note that the Stokes matrices $\ms{S}^{\pm}(q)$ can also be factorized
according to~\eqref{eq:Spn} in order to extract all the other Stokes
constants. This will be studied in detail in~\cite{GGM-peacock}.

We note that 
\begin{equation}
  \mathsf{S}^+_{\sigma_1 \sigma_1}(q)=  2H_{1,0}^+(q) H_{1,0}^- (q)
  =1 -12 q +3 q^2 +74 q^3 + 90 q^4+ O(q^5)
  = {\rm Ind}^{\rm rot}_{5_2}(q), 
\end{equation}
in agreement with Conjecture \ref{conj.2}.
 

\section{Open questions}
\label{sec.open}

In this paper we have formulated conjectures on the full resurgent
structure of quantum knot invariants of hyperbolic knots, and we have
presented detailed evidence for the first non-trivial cases, namely
the knots $4_1$ and $5_2$.  Although we used complex Chern-Simons
theory as a way to motivate our results, and state integrals and
asymptotic series as a way to present them, it is clear that a key
ingredient that controls the description of the asymptotic series in
Borel plane is a pair of linear $q$-difference equations with explicit
fundamental solutions. It is natural to ask whether these linear
$q$-difference equations are related to those that annihilate the
3D-index, or the colored Jones polynomial of a knot~\cite{GL}.  The
latter is the famous $\widehat{A}$-polynomial of a knot, whose
specialization at $q=1$ is conjectured to essentially coincide with
the $A$-polynomial of a knot~\cite{Ga:AJ}. It is an interesting
question to relate the newly found linear $q$-difference equations
with the $\widehat{A}$-polynomial of a knot.

One could also consider deformations by an arbitrary holonomy around
the knot, which will be explored in \cite{GGM-peacock}. In
this case, the resulting perturbative series depend on a parameter $x$
(see e.g. \cite{DGLZ}) that plays the role of a Jacobi variable and
one could calculate the Stokes constants in this extended
setting. This might make clearer the relation to the $A$-polynomial
and its quantization.

Another interesting question is whether the Stokes constants we
compute, which are closely related to BPS counting, can be obtained
with techniques similar to those of \cite{gmn2}, i.e. by doing WKB
analysis on the algebraic curve defined by the $A$-polynomial, or some
variant thereof.

Finally, we would like to point out that towers of singularities
similar to those studied here appear in the Borel plane of topological
string partition functions, see
e.g. \cite{pas-schiappa,cms}. Understanding the Stokes constants of
these singularities in topological string theory would probably lead
to fascinating mathematics and to connections with BPS state counting
in string theory.


\section*{Acknowledgements}

S.G. wishes to thank the Max Planck Institute for their hospitality
and especially Don Zagier. The authors would like to thank Jorgen Andersen,
Bertrand Eynard, Rinat Kashaev and Maxim Kontsevich for enlightening
conversations. The work of J.G. and M.M. is partially supported by the
Fonds National Suisse, subsidy 200020-175539, by the NCCR
51NF40-182902 ``The Mathematics of Physics'' (SwissMAP), and by the
ERC Synergy Grant ``ReNewQuantum".


\appendix

\section{$q$-series associated with the $5_2$ knot}
\label{sec.A52}

\subsection{The state integral of the $5_2$ knot}
\label{sub.52s}
  
  We now consider the case of $(A,B)=(2,3)$, i.e., the state integral
  $\calI_{2,3,\lambda,\mu}(\bb)$.
  The data we need to present our result are
  \begin{align}
    &F_{2,3,\lambda}(q,x) =
      \sum_{n=0}^\infty \frac{q^{n(n+1)+n\lambda}}{(q;q)_n^3}x^n,\\
    &\wt{F}_{2,3,\mu}(\tq,\tx) =
    \sum_{n=0}^\infty (-1)^n
    \frac{\tq^{\frac{1}{2}n(n+1)+n\mu}}{(\tq;\tq)_n^3}\tx^n,
  \end{align}
  as well as the operator
  \begin{align}
    P_{2,3,\lambda,\mu} =
    &-\frac{1}{2\pi\ri}+(1+2\delta-3\delta_1+\lambda)
      (\frac{1}{2}+\tilde{\delta} -3\tilde{\delta}_1 + \mu)\\
    &-\frac{\bb^2}{2}((1+2\delta-3\delta_1+\lambda)^2-3\delta_2)\\
    &-\frac{\bb^{-2}}{2}((\frac{1}{2}+\tilde{\delta} -3\tilde{\delta}_1
      + \mu)^2-\frac{1}{4}-3\tilde{\delta}_2 + 6\,
      E_2^{(0)}(\tq)).
  \end{align}
  Using the observation that
  \begin{equation}
    -\frac{1}{2\pi\ri} = \frac{1}{12}\left(\tau E_2(\tau) -
      \tau^{-1}E_2(-1/\tau)\right) 
  \end{equation}
  as well as
  \begin{equation}
    E_2(\tau) = 1-24\sum_{s=1}^\infty \frac{q^s}{(1-q^s)^2}
    = 1-24 E_2^{(0)}(q),
  \end{equation}
  the final result can be written as ($\tau = \bb^2$ in the upper half-plane)
  \begin{align}
    \calI_{2,3,\lambda,\mu}(\bb)
    &=
      (-1)^{\lambda-\mu}
      e^{\frac{\pi\ri}{4}}q^{\frac{\lambda}{2}}\tq^{\frac{\mu}{2}}
      \left(\frac{q}{\tq}\right)^{\frac{1}{8}}\times \\
    &\left(-\frac{\tau}{2}H_{2,\lambda}^{+}(q)H_{0,\mu}^{-}(\tq)
      +H_{1,\lambda}^{+}(q)H_{1,\mu}^{-}(\tq)
      -\frac{\tau^{-1}}{2}H_{0,\lambda}^{+}(q)H_{2,\mu}^{-}(\tq)\right).
      \label{eq:I23lm}
  \end{align}
  Here
  \begin{equation}
    H_{j,\lambda}^{+}(q) = \sum_{n=0}^\infty
    p_{n,\lambda}^{(j)}(q)t_{n,\lambda}(q),\quad
    H_{j,\mu}^{-}(\tq) = \sum_{n=0}^\infty
    P_{n,\mu}^{(j)}(\tq) T_{n,\mu}(\tq),\qquad (j=0,1,2) \,,
    \label{eq:gHpm}
  \end{equation}
  where $t_{n,\lambda}(q),T_{n,\mu}(\tq)$ are coefficients of
  $F_{2,3,\lambda}(q,x), \wt{F}_{2,3,\mu}(\tq,\tx)$ as series of
  $x,\tx$,
  \begin{align}
    &t_{n,\lambda}(q) = \frac{q^{n(n+1)+n\lambda}}{(q;q)_n^3}, \\
    &T_{n,\mu}(\tq) = (-1)^n\frac{\tq^{\frac{1}{2}n(n+1)+n\mu}}{(\tq;\tq)_n^3},
  \end{align}
  while $p_{n,\lambda}^{(j)}(q), P_{n,\mu}^{(j)}(\tq)$ result from
  applying the operator $P_{2,3,\lambda,\mu}$, i.e.
  \begin{align}
    &p_{n,\lambda}^{(0)}(q) = 1, \\ 
    &p_{n,\lambda}^{(1)}(q) = 1+2n-3E_1^{(n)}(q) + \lambda, \\ 
    &p_{n,\lambda}^{(2)}(q) = \left(p_{n,\lambda}^{(1)}(q)\right)^2
      - \frac{1}{6} - 3 E_2^{(n)}(q) + 4 E_2^{(0)}(q),
  \end{align}
  as well as
  \begin{align}
    &P_{n,\mu}^{(0)}(\tq) = 1, \\ 
    &P_{n,\mu}^{(1)}(\tq) = \frac{1}{2}+n-3E_1^{(n)}(\tq) + \mu, \\ 
    &P_{n,\mu}^{(2)}(\tq) = \left(P_{n,\mu}^{(1)}(\tq)\right)^2
      - \frac{1}{12} - 3 E_2^{(n)}(\tq) + 2 E_2^{(0)}(\tq).
  \end{align}
  The $q$-series $H_{j,\lambda}^{+}(q)$ and $H_{j,\mu}^{-}(q)$ for
  $j=0,1,2$ are exactly those that appear in Equation~\eqref{52fund1}.
  A few coefficients of the above series are given by
  \begin{equation}
    \begin{aligned}
      H_{0,-1}^{+}(q) =
      &1+q+3q^2+6q^3+11q^4+18q^5+\cO(q^6),\\
      H_{1,-1}^{+}(q) =
      &-q+3q^3+3q^4+3q^5+\cO(q^6),\\
      H_{2,-1}^{+}(q) =
      &-\frac{1}{6}+\frac{29}{6}q+\frac{55}{2}q^2+72q^3
      +\frac{895}{6}q^4+270q^5+\cO(q^6),\\
      H_{0,-1}^{-}(q) =
      &-2q-3q^2-2q^3+q^4+9q^5+\cO(q^6),\\
      H_{1,-1}^{-}(q) =
      &-1-3q-\frac{3}{2}q^2+12q^3+\frac{69}{2}q^4+\frac{153}{2}q^5+\cO(q^6),\\
      H_{2,-1}^{-}(q) =
      &\frac{5}{3}q+\frac{27}{2}q^2+\frac{143}{3}q^3+\frac{541}{6}q^4
      +\frac{263}{2}q^5+\cO(q^6).
    \end{aligned}
  \end{equation}

  \begin{equation}
    \begin{aligned}
      H_{0,0}^{+}(q) =
      &1+q^2+3q^3+6q^4+10q^5+\cO(q^6),\\
      H_{1,0}^{+}(q) =
      &1-3q-3q^2+3q^3+6q^4+12q^5+\cO(q^6),\\
      H_{2,0}^{+}(q) =
      &\frac{5}{6}-5q+\frac{53}{6}q^2+\frac{117}{2}q^3+117q^4
      +\frac{601}{3}q^5+\cO(q^6),\\
      H_{0,0}^{-}(q) =
      &1-q-3q^2-5q^3-7q^4-6q^5+\cO(q^6),\\
      H_{1,0}^{-}(q) =
      &\frac{1}{2}-\frac{9}{2}q-\frac{21}{2}q^2-\frac{19}{2}q^3
      -\frac{9}{2}q^4+27q^5+\cO(q^6),\\
      H_{2,0}^{-}(q) =
      &\frac{1}{6}-\frac{37}{6}q-\frac{17}{2}q^2+\frac{115}{6}q^3
      +\frac{389}{6}q^4+181q^5+\cO(q^6).
    \end{aligned}
  \end{equation}

  \begin{equation}
    \begin{aligned}
      H_{0,1}^{+}(q) =
      &1+q^3+3q^4+6q^5+\cO(q^6),\\
      H_{1,1}^{+}(q) =
      &2-3q-6q^2-2q^3+3q^4+15q^5+\cO(q^6),\\
      H_{2,1}^{+}(q) =
      &\frac{23}{6}-11q-12q^2+\frac{191}{6}q^3+\frac{189}{2}q^4
      +200q^5+\cO(q^6),\\
      H_{0,1}^{-}(q) =
      &1-q^2-3q^3-6q^4-9q^5+\cO(q^6),\\
      H_{1,1}^{-}(q) =
      &\frac{3}{2}-3q-\frac{17}{2}q^2-\frac{27}{2}q^3-21q^4
      -\frac{31}{2}q^5+\cO(q^6),\\
      H_{2,1}^{-}(q) =
      &\frac{13}{6}-10q-\frac{109}{6}q^2-\frac{13}{2}q^3+7q^4
      +\frac{173}{2}q^5+\cO(q^6).
    \end{aligned}
  \end{equation}
  

  \subsection{The symmetry $q \leftrightarrow q^{-1}$}
  \label{sub.qsym}

  We now discuss the symmetry $q \mapsto q^{-1}$. 
  It is easy to see that the above $q$-series $H_{j,\lambda}^{\pm}(q)$
  are well-defined holomorphic functions when $q$ is either inside or
  outside the unit disk. Extended this way, we claim that
  \begin{equation}
    \label{eq:Hpm-ac}
    H_{j,\lambda}^{+}(q) = (-1)^j H_{j,-\lambda}^{-}(1/q),
    \qquad (q \in \BC,\,\, |q| \neq 1) \,.
  \end{equation}

  This follows from the easy observation
  \be
  \label{tT}
    t_{n,\lambda}(q) = T_{n,-\lambda}(1/q) \,
\end{equation}
and the less trivial symmetry
\begin{equation}
\label{E12tau}
    E_1(\tau) = -E_1(-\tau),\quad E_2(\tau) = -E_2(-\tau)
\end{equation}
  (see for instance~\cite{bettin} for the first and~\cite{moller-zagier}
  for the second), where $E_1,E_2$ are related to $E_1^{(0)},E_2^{(0)}$ by
\begin{align}
E_1(\tau) &= 1 - 4\sum_{n\geq 1} \frac{q^n}{1-q^n} = 1- 4E_1^{(0)}(q), \\
E_2(\tau) &= 1 - 24\sum_{n\geq 1} \frac{q^n}{(1-q^n)^2} = 1-24E_2^{(0)}(q) \,.
\end{align}
Equations~\eqref{tT} and~\eqref{E12tau} and induction on the
exponent $n$ of $E_{1}^{(n)}, E_{2}^{(n)}$ imply that
  \begin{equation}
    p_{n,\lambda}^{(k)}(q) = (-1)^k P_{n,-\lambda}^{(k)}(1/q) 
  \end{equation}
  for all $n$, and this concludes the proof of Equation~\eqref{eq:Hpm-ac}.

  The above conclusions hold not only for the state integral with
  $(A,B)=(2,3)$ or $(A,B)=(1,2)$, but for the case of arbitrary
  integers $A$ and $B$ with $B > A > 0$.
  

\bibliographystyle{hamsalpha}
\bibliography{biblio}

\newcommand{\etalchar}[1]{$^{#1}$}
\providecommand{\bysame}{\leavevmode\hbox to3em{\hrulefill}\thinspace}
\providecommand{\href}[2]{#2}
\providecommand{\eprint}{\begingroup \urlstyle{rm}\Url}
\begin{thebibliography}{CMHR{\etalchar{+}}07}

\bibitem[ABS19]{abs}
In\^{e}s Aniceto, G\"{o}k\c{c}e Ba\c{s}ar, and Ricardo Schiappa, \emph{A primer
  on resurgent transseries and their asymptotics}, Phys. Rep. \textbf{809}
  (2019), 1--135.

\bibitem[AK14]{AK}
J{\o}rgen~Ellegaard Andersen and Rinat Kashaev, \emph{A {TQFT} from {Q}uantum
  {T}eichm\"uller theory}, Comm. Math. Phys. \textbf{330} (2014), no.~3,
  887--934.

\bibitem[BC13]{bettin}
Sandro Bettin and Brian Conrey, \emph{Period functions and cotangent sums},
  Algebra Number Theory \textbf{7} (2013), no.~1, 215--242.

\bibitem[BDP14]{Beem}
Christopher Beem, Tudor Dimofte, and Sara Pasquetti, \emph{Holomorphic blocks
  in three dimensions}, J. High Energy Phys. (2014), no.~12, 177, front
  matter+118.

\bibitem[CC01]{costin-costin}
O.~Costin and R.~D. Costin, \emph{On the formation of singularities of
  solutions of nonlinear differential systems in antistokes directions},
  Invent. Math. \textbf{145} (2001), no.~3, 425--485.

\bibitem[CG11]{CG}
Ovidiu Costin and Stavros Garoufalidis, \emph{Resurgence of the
  {K}ontsevich-{Z}agier series}, Ann. Inst. Fourier (Grenoble) \textbf{61}
  (2011), no.~3, 1225--1258.

\bibitem[CMHR{\etalchar{+}}07]{caliceti}
E.~Caliceti, M.~Meyer-Hermann, P.~Ribeca, A.~Surzhykov, and U.~D. Jentschura,
  \emph{From useful algorithms for slowly convergent series to physical
  predictions based on divergent perturbative expansions}, Phys. Rep.
  \textbf{446} (2007), no.~1-3, 1--96.

\bibitem[CMZ18]{moller-zagier}
Dawei Chen, Martin M\"{o}ller, and Don Zagier, \emph{Quasimodularity and large
  genus limits of {S}iegel-{V}eech constants}, J. Amer. Math. Soc. \textbf{31}
  (2018), no.~4, 1059--1163.

\bibitem[CSMnS17]{cms}
Ricardo Couso-Santamar\'{\i}a, Marcos Mari\~{n}o, and Ricardo Schiappa,
  \emph{Resurgence matches quantization}, J. Phys. A \textbf{50} (2017),
  no.~14, 145402, 34.

\bibitem[DG13]{DG}
Tudor Dimofte and Stavros Garoufalidis, \emph{The quantum content of the gluing
  equations}, Geom. Topol. \textbf{17} (2013), no.~3, 1253--1315.

\bibitem[DG18]{DG2}
\bysame, \emph{Quantum modularity and complex {C}hern-{S}imons theory}, Commun.
  Number Theory Phys. \textbf{12} (2018), no.~1, 1--52.

\bibitem[DGG13]{DGG2}
Tudor Dimofte, Davide Gaiotto, and Sergei Gukov, \emph{3-manifolds and 3d
  indices}, Adv. Theor. Math. Phys. \textbf{17} (2013), no.~5, 975--1076.

\bibitem[DGG14]{DGG1}
\bysame, \emph{Gauge theories labelled by three-manifolds}, Comm. Math. Phys.
  \textbf{325} (2014), no.~2, 367--419.

\bibitem[DGLZ09]{DGLZ}
Tudor Dimofte, Sergei Gukov, Jonatan Lenells, and Don Zagier, \emph{Exact
  results for perturbative {C}hern-{S}imons theory with complex gauge group},
  Commun. Number Theory Phys. \textbf{3} (2009), no.~2, 363--443.

\bibitem[Fad95]{Faddeev}
L.~D. Faddeev, \emph{Discrete {H}eisenberg-{W}eyl group and modular group},
  Lett. Math. Phys. \textbf{34} (1995), no.~3, 249--254.

\bibitem[FK94]{FK-QDL}
Ludwig Faddeev and Rinat Kashaev, \emph{Quantum dilogarithm}, Modern Phys.
  Lett. A \textbf{9} (1994), no.~5, 427--434.

\bibitem[Gar04]{Ga:AJ}
Stavros Garoufalidis, \emph{On the characteristic and deformation varieties of
  a knot}, Proceedings of the {C}asson {F}est, Geom. Topol. Monogr., vol.~7,
  Geom. Topol. Publ., Coventry, 2004, pp.~291--309 (electronic).

\bibitem[Gar08]{Ga:resurgence}
Stavros Garoufalidis, \emph{Chern-{S}imons theory, analytic continuation and
  arithmetic}, Acta Math. Vietnam. \textbf{33} (2008), no.~3, 335--362.

\bibitem[GGMn20a]{GGM-peacock}
Stavros Garoufalidis, Jie Gu, and Marcos Mari\~{n}o, \emph{{Peacock patterns
  and resurgence in complex Chern-Simons theory}}, \eprint{2012.00062}.

\bibitem[GGMn20b]{ggm}
Alba Grassi, Jie Gu, and Marcos Mari\~no, \emph{{Non-perturbative approaches to
  the quantum Seiberg-Witten curve}}, JHEP \textbf{07} (2020), 106,
  \eprint{1908.07065}.

\bibitem[GH18]{gh-res}
Dongmin Gang and Yasuyuki Hatsuda, \emph{S-duality resurgence in {$\rm SL(2)$}
  {C}hern-{S}imons theory}, J. High Energy Phys. (2018), no.~7, 053, front
  matter+23.

\bibitem[GK17]{GK:qseries}
Stavros Garoufalidis and Rinat Kashaev, \emph{From state integrals to
  {$q$}-series}, Math. Res. Lett. \textbf{24} (2017), no.~3, 781--801.

\bibitem[GL05]{GL}
Stavros Garoufalidis and Thang T.~Q. L{\^e}, \emph{The colored {J}ones function
  is {$q$}-holonomic}, Geom. Topol. \textbf{9} (2005), 1253--1293 (electronic).

\bibitem[GLMn08]{GLM}
Stavros Garoufalidis, Thang T.~Q. L\^{e}, and Marcos Mari\~{n}o,
  \emph{Analyticity of the free energy of a closed 3-manifold}, SIGMA Symmetry
  Integrability Geom. Methods Appl. \textbf{4} (2008), Paper 080, 20.

\bibitem[GMN10]{gmn}
Davide Gaiotto, Gregory~W. Moore, and Andrew Neitzke, \emph{Four-dimensional
  wall-crossing via three-dimensional field theory}, Comm. Math. Phys.
  \textbf{299} (2010), no.~1, 163--224.

\bibitem[GMN13]{gmn2}
\bysame, \emph{Wall-crossing, {H}itchin systems, and the {WKB} approximation},
  Adv. Math. \textbf{234} (2013), 239--403.

\bibitem[GMnP]{gmp}
Sergei Gukov, Marcos Mari\~no, and Pavel Putrov, \emph{Resurgence in complex
  {C}hern-{S}imons theory}, \eprint{arXiv:1605.07615}, Preprint 2016.

\bibitem[Guk05]{gukov}
Sergei Gukov, \emph{Three-dimensional quantum gravity, {C}hern-{S}imons theory,
  and the {A}-polynomial}, Comm. Math. Phys. \textbf{255} (2005), no.~3,
  577--627.

\bibitem[GZa]{GZ:asymptotics}
Stavros Garoufalidis and Don Zagier, \emph{Asymptotics of {N}ahm sums at roots
  of unity}, With an appendix by Sander Zwegers.

\bibitem[GZb]{GZ:qseries}
\bysame, \emph{Knots and their related $q$-series}, Preprint 2020.

\bibitem[GZc]{GZ:kashaev}
\bysame, \emph{Knots, perturbative series and quantum modularity}, Preprint
  2020.

\bibitem[Hik07]{Hikami}
Kazuhiro Hikami, \emph{Generalized volume conjecture and the {$A$}-polynomials:
  the {N}eumann-{Z}agier potential function as a classical limit of the
  partition function}, J. Geom. Phys. \textbf{57} (2007), no.~9, 1895--1940.

\bibitem[IMnS19]{ims}
Katsushi Ito, Marcos Mari\~{n}o, and Hongfei Shu, \emph{T{BA} equations and
  resurgent quantum mechanics}, J. High Energy Phys. (2019), no.~1, 228, front
  matter+44.

\bibitem[Kas95]{K95}
Rinat Kashaev, \emph{A link invariant from quantum dilogarithm}, Modern Phys.
  Lett. A \textbf{10} (1995), no.~19, 1409--1418.

\bibitem[Kas97]{kas-volume}
R.~M. Kashaev, \emph{The hyperbolic volume of knots from the quantum
  dilogarithm}, Lett. Math. Phys. \textbf{39} (1997), no.~3, 269--275.

\bibitem[Kon20]{kontsevich-talk}
Maxim Kontsevich, \emph{{Exponential integrals, Lefschetz thimbles and linear
  resurgence}}, {June 2020}, ReNewQuantum seminar,
  \url{https://renewquantum.eu/docs/Lecture_slides_MaximKontsevich_June2020.pdf},
  2020.

\bibitem[KS11]{KS}
Maxim Kontsevich and Yan Soibelman, \emph{Cohomological {H}all algebra,
  exponential {H}odge structures and motivic {D}onaldson-{T}homas invariants},
  Commun. Number Theory Phys. \textbf{5} (2011), no.~2, 231--352.

\bibitem[KS20]{ks-ar}
\bysame, \emph{{Analyticity and resurgence in wall-crossing formulas}},
  \eprint{2005.10651}.

\bibitem[Mn19]{mm-s2019}
Marcos Mari\~no, \emph{{From resurgence to BPS states}}, {talk given at the
  conference {\it Strings 2019}}, Brussels,
  \url{https://livestream.com/accounts/7777696/events/8742238/videos/193704304},
  2019.

\bibitem[MS16]{msauzin}
Claude Mitschi and David Sauzin, \emph{Divergent series, summability and
  resurgence. {I}}, Lecture Notes in Mathematics, vol. 2153, Springer, 2016.

\bibitem[PS10]{pas-schiappa}
Sara Pasquetti and Ricardo Schiappa, \emph{Borel and {S}tokes nonperturbative
  phenomena in topological string theory and {$c=1$} matrix models}, Ann. Henri
  Poincar\'{e} \textbf{11} (2010), no.~3, 351--431.

\bibitem[Vor83]{voros}
A.~Voros, \emph{The return of the quartic oscillator: the complex {WKB}
  method}, Ann. Inst. H. Poincar\'{e} Sect. A (N.S.) \textbf{39} (1983), no.~3,
  211--338.

\end{thebibliography}
\end{document}